\documentclass[onecolumn,aps,prd,superscriptaddress,floatfix,notitlepage,abstract]{revtex4-2}
\usepackage{graphicx,amssymb,amsmath,epsfig,mathtools, caption,csquotes} % Required for inserting images
\usepackage{palatino}
\usepackage{natbib}
\usepackage{tabularx}
\usepackage{booktabs}
\usepackage{multirow}
\usepackage{array}
\usepackage{overpic}
\usepackage{subcaption}
\usepackage{placeins} 
\usepackage{amsthm}
\usepackage{booktabs}
\usepackage[hidelinks]{hyperref}
\newcommand{\VA}[1]{\parbox[t]{0.44\textwidth}{\raggedright #1}}
\newcommand{\VB}[1]{\parbox[t]{0.23\textwidth}{\raggedright #1}}
\newcommand{\VC}[1]{\parbox[t]{0.22\textwidth}{\raggedright #1}}
\newtheorem{theorem}{Theorem}
\newtheorem{lemma}{Lemma}
\newtheorem{proposition}{Proposition}
\newtheorem{corollary}{Corollary}
\theoremstyle{definition}
\newtheorem{definition}{Definition}
\theoremstyle{remark}
\newtheorem{remark}{Remark}

\newcommand{\Teff}{T_{\rm eff}}
\newcommand{\Peff}{P_{\rm eff}}
\newcommand{\dd}{\delta}

\begin{document}
\title{Universality and calibration dependence of effective two-horizon thermodynamics in de~Sitter spacetimes}
\author{Anees Ashfaq}
\email{aneesashfaq2002@gmail.com}
\affiliation{Department of Mathematics, Faculty of Science, University of Okara, Okara 56130, Pakistan.}
\author{Nazek Alessa}
\affiliation{Department of Mathematical Sciences, College of Science,
Princess Nourah bint Abdulrahman University, P.O.Box 84428, Riyadh
11671, Saudi Arabia}

\author{M. Umair Shahzad}
\email{mushahzad@uo.edu.pk;m.u.shahzad@ucp.edu.pk}
\affiliation{Department of Mathematics, Faculty of Science, University of Okara, Okara 56130, Pakistan.}
\affiliation{Department of Technical Sciences, Western Caspian University, Baku, Azerbaijan}
\begin{abstract}
The effective two-horizon method describes a de~Sitter black hole and its cosmological horizon as one thermodynamic system, with entropy $S=\pi r_c^2F(x)$, $x=r_+/r_c$, where $F$ is fixed by a calibration condition. For the class of spacetimes $f(r)=k-2M/r-\Lambda r^2/3+H(r)$ with arbitrary $H$, we show that the calibration equation for $F$ does not depend on $H$ or $k$, so that $F$ is the same for every member, and that the effective temperature always has the form $\Teff=\Omega(x)(T_c+T_+/x^4)$, with $\Omega$ fixed by the calibration alone. This universality is a consistency property of the entropy ansatz, and it replaces a series of case-by-case derivations by one statement. It holds for every normalization of temperature, energy and entropy in which each horizon obeys its own first law, $dE=T_+dS_+$ at the black-hole horizon and $dE=-T_c\,dS_c$ at the cosmological horizon; with mixed normalizations, such as the Bousso--Hawking temperature together with the mass as energy, $F$ depends on the metric. The cold and lukewarm calibrations found in the literature are incompatible for every member of the class. We derive an exact zero condition for the canonical heat capacity, whose root lies within $0.3\%$ of a universal value in the ten configurations examined, with a deviation that grows with the charge. We also derive the small-black-hole and Nariai limits.
\end{abstract}
\maketitle
\section{Introduction}
\label{sec:intro}

A black hole in de Sitter space has two horizons, the black-hole horizon $r_+$ and the cosmological horizon $r_c$, which in general have different temperatures $T_+$ and $T_c$~\cite{GH1977}. The region between them is therefore not in thermal equilibrium, and the standard thermodynamics of a single horizon does not apply directly. Several approaches have been developed to deal with this, including separate first laws for each horizon, the mechanical first law, and the inclusion of $\Lambda$ as a thermodynamic variable~\cite{Cai2002, Sekiwa2006,Urano2009, Dolan2013,KS2016, KMT2017}.

A separate line of work treats the thermodynamics of de~Sitter black holes from first principles. Gomberoff and Teitelboim showed that either horizon can be used as the boundary that fixes the ensemble, and that the two resulting systems have energies of equal magnitude and opposite sign~\cite{GT2003}. Banihashemi and Jacobson constructed canonical and microcanonical ensembles with cosmological horizons by introducing a boundary~\cite{BJ2022}, and the minus sign in the first law of the cosmological horizon was analysed in Ref.~\cite{BJSV2023}. The Euclidean action of Schwarzschild--de~Sitter black holes, evaluated with constrained instantons, is minus the sum of the two horizon entropies~\cite{MvdSV2023}, and the same holds for charged black holes in de~Sitter space for arbitrary mass and charge~\cite{MvdSV2024}. In the variables used below, this corresponds to $S=\pi r_c^2(1+x^2)$. The effective method studied in this paper does not derive the entropy from an action. It fixes it by a calibration condition, and our results characterize what follows from that choice.

One approach to this problem treats the two horizons as a single effective system. The entropy is written as $S=\pi r_c^2F(x)$ with $x=r_+/r_c$, the volume of the region between the horizons is taken as the thermodynamic volume, and an effective temperature $\Teff$ is defined through a first law $dM=\Teff\,dS-\Peff\,dV+\cdots$. The function $F(x)$, which encodes the correlation between the horizons, is not fixed by the first law alone. It is fixed by a calibration condition: a requirement that $\Teff$ reduces to a known temperature in a particular configuration.

Effective quantities for the two-horizon system were introduced earlier for RN--dS with $S=\pi r_c^2(1+x^2)$~\cite{Ma2014}; see also Ref.~\cite{LiMaMa2017}. The entropy function was introduced for Reissner--Nordstr\"om--de~Sitter (RN--dS) spacetime by Zhang, Zhao and Ma~\cite{ZZM2016}, who fixed $F$ by requiring $\Teff=T_+=T_c$ in the lukewarm configuration. 
Ma, Zhao and Cao~\cite{MZC2018} applied it to de~Sitter black holes in massive gravity, found the same function, and argued that it should be universal because the entropy depends only on the horizon positions. The lukewarm calibration was used again in Ref.~\cite{DLZ2022} and remains in use~\cite{Sereewat2026}; the method has also been extended to higher dimensions~\cite{ZhaoZhang2018}. A different calibration, requiring $\Teff=T_c$ when the black-hole horizon is cold ($T_+=0$), was introduced in Ref.~\cite{Zhen2025} and has since been applied to a number of spacetimes~\cite{KRdS2024,MaZhao2025,Paper1,Paper2,Paper3}. In each case $F(x)$ is derived a new metric for the system at hand, and the effective quantities are then used to study heat capacities, Schottky-like peaks and phase structure. Schottky-like peaks in black-hole heat capacities, including for de Sitter black holes, have also been studied independently of this method~\cite{Johnson2020,JohnsondS,Dinsmore2020}.

The same entropy function has in fact been found repeatedly, case by case. With the lukewarm calibration, the function of Refs.~\cite{MZC2018,DLZ2022} was obtained again for charged de~Sitter black holes with a cloud of strings and quintessence~\cite{Ma2019CSQ}, for RN--dS spacetime with quintessence~\cite{Zhang2022Q}, and for de~Sitter spacetime with a Kalb--Ramond field~\cite{Du2024KR}. In each of these works, it was derived again for the metric at hand, and the agreement with RN--dS was not explained. Effective quantities of this type were also used to study the Schottky anomaly of a nonlinearly charged dS spacetime~\cite{Zhen2024CQG}. For the charged Gauss--Bonnet black hole in dS space, by contrast, the entropy needs a second function multiplied by the Gauss--Bonnet coupling~\cite{Guo2020GB}. 
The universality shown below is a consistency property of the entropy ansatz~\eqref{eq:SV}: once the horizon conditions $f(r_\pm)=0$ are written as a linear system in $M$ and $\Lambda$, the calibration equation contains no further information about the metric. What is new is the statement for a whole class of spacetimes with a precise boundary, the condition $as=b$ on the normalization of temperature, energy, and entropy under which it holds, the incompatibility of the two calibrations for every member of the class, and a quantitative test of how published heat-capacity peaks change when the calibration is changed.

Two questions have not been addressed. First, what does $F(x)$ depend on? Each application re-derives it, and the argument of Ref.~\cite{MZC2018} concerns one calibration and one example. Second, how are the two calibrations related? Ref.~\cite{Zhen2025} notes that its effective quantities differ from those of Ref.~\cite{DLZ2022} because of the different boundary conditions, but the two choices have not been compared.

In this paper, we answer both questions for the class of static, spherically symmetric spacetime $f(r)=k-2M/r-\Lambda r^2/3+H(r)$, with arbitrary $H(r)$ and constant $k$. Our main results are the following.
\begin{enumerate}
\item \emph{Universality.} For every spacetime in the class, the calibration ODE for $F(x)$ is independent of $H$ and $k$. With the cold calibration and $F(0)=1$, its unique solution is $F(x)=\tfrac85(1-x^3)^{2/3}+(x^3-\tfrac35)/(1-x^3)$. The same holds for the lukewarm calibration, with a different $F$ (Theorem~\ref{thm:universal}).
\item \emph{Structure of $\Teff$.} For every spacetime in the class, $\Teff=\Omega(x)\left(T_c+T_+/x^4\right)$, where $\Omega$ depends only on the calibration. The cold calibration gives $\Teff=T_c+T_+/x^4$ exactly; the metric enters only through the two horizon temperatures (Theorem~\ref{thm:structure}).
\item \emph{Normalization.} Universality holds for any normalization of temperature, energy, and entropy in which each horizon separately obeys  $dE=T_+dS_+$ at $r_+$ and $dE=-T_c\,dS_c$ at $r_c$~\cite{GH1977, BJSV2023}, and fails otherwise. In particular, it fails for the Bousso--Hawking normalization of the temperature with the mass as energy (Section~\ref{sec:normalization}).
\item \emph{Incompatibility.} The cold and lukewarm calibrations cannot both hold for any spacetime in the class. They lead to different entropy functions, and hence to different heat capacities and limiting behaviour (Corollary~\ref{cor:conflict} and Section~\ref{sec:calibration}).
\item \emph{Consequence.}  The Schottky-like maxima of the canonical heat capacity, and the microstate counting matched to them, depend on the calibration. We show this for the published spacetimes of Refs.~\cite{Zhen2025,KRdS2024,Paper1,Paper2,Paper3}: under the other calibration, the maximum moves to a different horizon ratio, or into a narrow window next to the cold boundary. (Section~\ref{sec:paper3luke}.)
\end{enumerate}
We also state the boundary of the class, with a counterexample for each of the assumptions \(A1\), \(A2\) and \(A4\) (\(A3\) defines the setting), and derive general results (Smarr relation, heat capacity, limits) that follow from these theorems. These results show what the method depends on. They do not by themselves establish that the method, or either calibration, is physically correct.

The paper is organized as follows. Section~\ref{sec:class} defines the class and the effective quantities. Section~\ref{sec:theorem} proves the universality theorem and states the class boundary, including the role of normalization. Section~\ref{sec:structure} derives the structure of $\Teff$ and the incompatibility of the calibrations. Section~\ref{sec:general} gives general results. Section~\ref{sec:examples} applies them to four spacetimes, and Section~\ref{sec:calibration} shows which conclusions change with the calibration. Section~\ref{sec:conclusions} concludes. Appendix~\ref{app:verification} lists the symbolic and numerical checks.

\section{Metric class and effective quantities}
\label{sec:class}

\subsection{The class}

We use units $G=c=\hbar=k_B=1$ throughout.
We consider static, spherically symmetric spacetimes
\begin{equation}
ds^2=-f(r)\,dt^2+\frac{dr^2}{f(r)}+r^2d\Omega^2,
\qquad
f(r)=k-\frac{2M}{r}-\frac{\Lambda r^2}{3}+H(r),
\label{eq:metric}
\end{equation}
where $d\Omega_2^2$ is the metric of the unit two-sphere, $M$ is the mass parameter, $\Lambda>0$ the cosmological constant, $k$ a constant and $H(r)$ a function collecting all further matter or modified-gravity contributions. Where a length scale $\ell$ is used, it is defined by $\Lambda=3/\ell^2$. We denote by $\{q_i\}$ the parameters contained in $k$ and $H$ (charges, couplings). We assume:
\begin{itemize}
\item[\textbf{(A1)}] $k$ and $H(r)$ do not depend on $M$ or $\Lambda$;
\item[\textbf{(A2)}] the same $f(r)$ appears in $g_{tt}$ and $g_{rr}$, and temperature, energy and entropy are normalized as in Eq.~\eqref{eq:temps} and the text below it (relaxed in Section~\ref{sec:normalization});
\item[\textbf{(A3)}] $f$ has a black-hole horizon $r_+$ and a cosmological horizon $r_c>r_+$;
\item[\textbf{(A4)}] the entropy and volume of the effective system are given by Eq.~\eqref{eq:SV} below.
\end{itemize}
We write $x\equiv r_+/r_c\in(0,1)$, $u\equiv1-x^3$ and $A\equiv1+x+x^2$.

\subsection{Horizon conditions and temperatures}

The conditions $f(r_+)=f(r_c)=0$ are linear in $M$ and $\Lambda$. Solving them gives, for any $H$,
\begin{align}
M(r_c,x)&=\frac{x\,r_c\left[k+H(xr_c)-x^2\big(k+H(r_c)\big)\right]}{2\left(1-x^3\right)}, \label{eq:Mrcx}\\
\Lambda(r_c,x)&=\frac{3\left[k+H(r_c)-x\big(k+H(xr_c)\big)\right]}{r_c^2\left(1-x^3\right)} . \label{eq:Lrcx}
\end{align}
Subtracting $rf(r)=0$ at the two horizons gives the equivalent relation
\begin{equation}
\frac{\Lambda}{3}\,r_c^2A=k+\frac{r_cH(r_c)-r_+H(r_+)}{r_c(1-x)} ,
\label{eq:rcLambda}
\end{equation}
which fixes $r_c(x)$ at fixed $\Lambda$ and $\{q_i\}$. For RN--dS ($k=1$, $H=Q^2/r^2$) the last term equals $-Q^2/(xr_c^2)$.

The horizon temperatures are
\begin{equation}
T_+=\frac{f'(r_+)}{4\pi},\qquad T_c=-\frac{f'(r_c)}{4\pi},
\label{eq:temps}
\end{equation}
where the sign at $r_c$ follows from $f'(r_c)<0$. They are $T_h=\kappa_h/2\pi$, where $\kappa_h$ is the surface gravity of the Killing vector $\xi=\partial_t$ normalized as in Eq.~\eqref{eq:metric}, so that $\xi^2=-f$. The system's energy is the mass parameter $M$, and the entropy of each horizon is a quarter of its area. This is the normalization used in the applications of the method~\cite{ZZM2016,MZC2018,DLZ2022,Zhen2025,KRdS2024,Paper2,Paper3}. Other normalizations are treated in Section~\ref{sec:normalization}.

\subsection{Effective quantities}

\begin{definition}
\label{def:effective}
The entropy and volume of the effective system are
\begin{equation}
S=\pi r_c^2F(x),\qquad V=\frac{4\pi}{3}\left(r_c^3-r_+^3\right),
\label{eq:SV}
\end{equation}
with $F$ a function of $x$ only. The effective temperature, pressure, and coupling potentials are defined by the first law
\begin{equation}
dM=\Teff\,dS-\Peff\,dV+\sum_i\varphi_i\,dq_i ,
\label{eq:firstlaw}
\end{equation}
that is,
\begin{equation}
\Teff=\left(\frac{\partial M}{\partial S}\right)_{V,\{q\}},\qquad
\Peff=-\left(\frac{\partial M}{\partial V}\right)_{S,\{q\}},\qquad
\varphi_i=\left(\frac{\partial M}{\partial q_i}\right)_{S,V} .
\label{eq:Teffdef}
\end{equation}
\end{definition}

Since $S$ and $V$ depend only on $r_+$ and $r_c$, the potentials follow directly from Eq.~\eqref{eq:Mrcx}:
\begin{equation}
\varphi_i=\frac{x\,r_c}{2(1-x^3)}\left[\frac{\partial\big(k+H(r_+)\big)}{\partial q_i}-x^2\,\frac{\partial\big(k+H(r_c)\big)}{\partial q_i}\right].
\label{eq:potentials}
\end{equation}
It is convenient to define
\begin{equation}
G(x)\equiv(1-x^3)F'(x)+2x^2F(x).
\label{eq:G}
\end{equation}
$\Teff$ and $\Peff$ are given for the whole class in Lemma~\ref{lem:dM} and Eq.~\eqref{eq:Peff}.

\section{Universality theorem} \label{sec:theorem}

\subsection{The key lemma}

\begin{lemma}
\label{lem:dM}
For every spacetime satisfying (A1)--(A4),
\begin{equation}
\Teff=\frac{2x\left(T_++x^4T_c\right)}{(1-x^3)\,G(x)} .
\label{eq:Teffgeneral}
\end{equation}
\end{lemma}

\begin{proof}
Consider a small change at fixed $V$ and $\{q_i\}$, in which $M$ and $\Lambda$ vary. Varying $f(r_+)=0$ gives
\begin{equation}
f'(r_+)\,\dd r_+-\frac{2}{r_+}\,\dd M-\frac{r_+^2}{3}\,\dd\Lambda=0 .
\end{equation}
The term $f'(r_+)\,\dd r_+$ contains every contribution from the change of $r_+$, including $H'(r_+)\,\dd r_+$. By (A1), $H$ and $k$ give no terms in $\dd M$ or $\dd\Lambda$. With Eq.~\eqref{eq:temps}, and in the same way at $r_c$,
\begin{align}
4\pi T_+\,\dd r_+&=\frac{2}{r_+}\,\dd M+\frac{r_+^2}{3}\,\dd\Lambda, \label{eq:var1}\\
-4\pi T_c\,\dd r_c&=\frac{2}{r_c}\,\dd M+\frac{r_c^2}{3}\,\dd\Lambda. \label{eq:var2}
\end{align}
Multiplying Eq.~\eqref{eq:var1} by $r_c^2$ and Eq.~\eqref{eq:var2} by $r_+^2$ and subtracting eliminates $\dd\Lambda$:
\begin{equation}
\dd M=\frac{2\pi r_+r_c\left(r_c^2T_+\,\dd r_++r_+^2T_c\,\dd r_c\right)}{r_c^3-r_+^3}.
\label{eq:dM1}
\end{equation}
At fixed volume $\dd r_c=x^2\,\dd r_+$, so that
\begin{equation}
\dd M=\frac{2\pi x\,r_c\left(T_++x^4T_c\right)}{1-x^3}\,\dd r_+ .
\label{eq:dM}
\end{equation}
Along the same path $\dd x=(1-x^3)\,\dd r_+/r_c$, and Eq.~\eqref{eq:SV} gives
\begin{equation}
\dd S=\pi\left(2r_cF\,\dd r_c+r_c^2F'\,\dd x\right)=\pi r_c\,G(x)\,\dd r_+ .
\label{eq:dS}
\end{equation}
Dividing Eq.~\eqref{eq:dM} by Eq.~\eqref{eq:dS} gives Eq.~\eqref{eq:Teffgeneral}.
\end{proof}

The same method applied to Eq.~\eqref{eq:dM1} with both radii free gives the effective pressure,
\begin{equation}
\Peff=\frac{x\left[2T_+F-xF'\left(T_++xT_c\right)\right]}{2r_c\,(1-x^3)\,G(x)} ,
\label{eq:Peff}
\end{equation}
which is also of the same form for every spacetime in the class.

\subsection{The calibration ODE}

\begin{theorem}[Universality]
\label{thm:universal}
Suppose that for every $x\in(0,1)$ the spacetime admits a state with $T_+=0$ and $T_c\neq0$. Then the cold calibration
\begin{equation}
\Teff\big|_{T_+=0}=T_c
\label{eq:coldcond}
\end{equation}
is equivalent to
\begin{equation}
(1-x^3)F'+2x^2F=\frac{2x^5}{1-x^3},
\label{eq:coldODE}
\end{equation}
for every spacetime in the class. With $F(0)=1$, its unique solution on $[0,1)$ is
\begin{equation}
F_{\rm cold}(x)=\frac85\left(1-x^3\right)^{2/3}+\frac{x^3-\frac35}{1-x^3} .
\label{eq:Fcold}
\end{equation}
\end{theorem}

\begin{proof}
At a state with $T_+=0$, Lemma~\ref{lem:dM} gives $\Teff=2x^5T_c/[(1-x^3)G]$. Setting this equal to $T_c$ and dividing by $T_c\neq0$ gives Eq.~\eqref{eq:coldODE} at that value of $x$. Since such a state exists for every $x$, Eq.~\eqref{eq:coldODE} holds on $(0,1)$. Neither $H$ nor $k$ appears in it.

\emph{Existence and uniqueness.} Eq.~\eqref{eq:coldODE} is linear and first order, $F'+p(x)F=s(x)$, with $p=2x^2/u$ and $s=2x^5/u^2$ continuous on $[0,1)$. By the standard theorem for linear ODEs, the solution with $F(0)=1$ exists and is unique on $[0,1)$.

\emph{Solution.} The integrating factor is $u^{-2/3}$, so $\tfrac{d}{dx}\big[F\,u^{-2/3}\big]=2x^5u^{-8/3}$. With $x^5dx=-\tfrac13(1-u)\,du$, the integral is $\tfrac25u^{-5/3}-u^{-2/3}+C$, so $F=\tfrac{2}{5u}-1+Cu^{2/3}$. The condition $F(0)=1$, under which the entropy reduces to that of the cosmological horizon at $x=0$, gives $C=8/5$ and Eq.~\eqref{eq:Fcold}.
\end{proof}

\paragraph{Reaching the calibration states.} At a given $x$, a cold state is reached by choosing $\Lambda$ and the $\{q_i\}$, together with $M$, so that $f(r_+)=f'(r_+)=f(r_c)=0$; a lukewarm state is reached in the same way with $f'(r_+)=-f'(r_c)$ in place of $f'(r_+)=0$. For RN--dS, both lines exist for every $x\in(0,1)$. In units of $r_c$, the cold line is
\begin{equation}
M=\frac{x(1+x)^2}{1+2x+3x^2},\quad Q^2=\frac{x^2(1+2x)}{1+2x+3x^2},\quad \Lambda=\frac{3}{1+2x+3x^2},
\label{eq:coldline}
\end{equation}
which is positive on $(0,1)$ and runs from pure de~Sitter space at $x=0$ to the ultracold point $M=2/3$, $Q^2=\Lambda=1/2$ at $x=1$; the lukewarm line is $M=Q=x/(1+x)$, $\Lambda=3/(1+x)^2$. The same holds for the nonlinear-electrodynamics example of Section~\ref{sec:examples}, in which $k$ only rescales these values, and for Conformal Killing gravity at fixed $\lambda r_c^4=0.1$, we checked numerically, at $x=0.3$, $0.5$ and $0.7$ only, that $Q^2$ and $\Lambda$ on the cold line are positive (Appendix~\ref{app:verification}).

For Conformal Killing gravity, Eq.~\eqref{eq:coldODE} is the equation derived in Ref.~\cite{Paper2}. Theorem~\ref{thm:universal} shows that it is the same equation for every member of the class, and Eq.~\eqref{eq:Fcold} need not be re-derived for each new metric.

\subsection{Class boundary}
\label{sec:boundary}

Assumptions \(A1\), \(A2\) and \(A4\) each enter the proof at one point, and breaking any of them makes $F$ depend on a parameter of the spacetime; \(A3\) only defines the setting. If the hypothesis of Theorem~\ref{thm:universal} fails, the calibration cannot be applied (last paragraph of this subsection).

\paragraph{\(A1\) $H$ depends on $M$.} For $f=1-2M/r+Q^2/r^2+\beta M^2/r^2-\Lambda r^2/3$, varying $\beta M^2/r^2$ gives an extra term $(2\beta M/r^2)\,\dd M$ in Eqs.~\eqref{eq:var1}--\eqref{eq:var2}. At $(r_+,r_c,Q,\beta)=(1,3,0.4,0.1)$, on the branch continuously connected to RN--dS ($M\simeq0.5597$), a direct computation gives $\Teff/(T_c+T_+/x^4)=1.061$, a deviation of $6\%$. %REV3 minor

\paragraph{Mass term not $-2M/r$.} The metric $f=k-2(1+\gamma)M/r-\Lambda r^2/3+H$ is a member of the class with the mass parameter relabelled, $M\to(1+\gamma)M$; the point is that the energy is still taken to be $M$. The same steps give $\Teff=\Omega(x)(T_c+T_+/x^4)/(1+\gamma)$. The cold calibration then fixes $G=2x^5/[(1+\gamma)(1-x^3)]$, so $F$ depends on $\gamma$. This is the case $a=1/(1+\gamma)$ of Eq.~\eqref{eq:Teffnorm} below.

\paragraph{\(A4\) Non-area entropy.} Generalized entropies such as those of Barrow, Tsallis and R\'enyi~\cite{Barrow2020,TC2013,CI2016} change the scaling of $S$. For $S=c\,r_c^{2+\delta}F(x)$ with $\delta\neq0$, Eq.~\eqref{eq:dS} becomes $\dd S=c\,r_c^{1+\delta}\left[(1-x^3)F'+(2+\delta)x^2F\right]\dd r_+$, while Eq.~\eqref{eq:dM} is unchanged. Then $\Teff\propto r_c^{-\delta}$, and no function $F(x)$ of $x$ alone can satisfy the calibration. The same occurs in $D\neq4$ dimensions, where $S\propto r_c^{D-2}$ and $V\propto r^{D-1}$~\cite{ZhaoZhang2018,Hairy5D}. A further example is the charged Gauss--Bonnet black hole in dS space~\cite{Guo2020GB}, whose horizon entropy contains, besides the area term, a term proportional to the Gauss--Bonnet coupling $\alpha$ with a different power of the radius. The effective entropy then takes the form $S\propto r_c^{3}\big[f(x)+(\alpha/r_c^2)f_1(x)\big]$, with a second function $f_1$, and its dependence on $x$ cannot be reduced to a single function fixed by the calibration.

\paragraph{No calibration state.} For Schwarzschild--de~Sitter ($H=0$) there is no state with $T_+=0$ inside the physical domain, so the cold calibration cannot be applied to it directly. Since Eq.~\eqref{eq:coldODE} is the same for every member of the class, $F$ may be defined for such spacetimes by continuity from members that admit a cold state, but this is a choice and not a consequence of the calibration.

\subsection{Normalization of temperature, energy and entropy}
\label{sec:normalization}

Assumption (A2) fixes three normalizations: of the temperature (Killing vector $\partial_t$ with $\xi^2=-f$), of the energy ($E=M$), and of the entropy (a quarter of the area). Suppose instead that $f$ is still of the form~\eqref{eq:metric}, but
\begin{equation}
T_h=\frac{|f'(r_h)|}{4\pi s},\qquad E=aM,\qquad S=b\,\pi r_c^2F(x),
\label{eq:norm}
\end{equation}
with positive constants $a$, $b$, $s$, and that $\Teff=(\partial E/\partial S)_{V,\{q\}}$. Repeating the proof of Lemma~\ref{lem:dM} with $\dd E=a\,\dd M$ gives
\begin{equation}
\Teff=\frac{as}{b}\,\frac{2x\left(T_++x^4T_c\right)}{(1-x^3)\,G(x)} ,
\label{eq:Teffnorm}
\end{equation}
so the cold calibration fixes $G=(as/b)\,2x^5/(1-x^3)$, and $F$ depends on the normalization unless $as=b$. The condition $as=b$ has a direct meaning. For a single horizon at fixed $\Lambda$ and $\{q_i\}$, with $S_h=b\pi r_h^2$ and $M(r_h)$ from $f(r_h)=0$,
\begin{equation}
\frac{dE}{dr_h}-\frac{f'(r_h)}{4\pi s}\,\frac{dS_h}{dr_h}=\frac{as-b}{2s}\left(k+H+r_hH'-\Lambda r_h^2\right),
\label{eq:singlefirstlaw}
\end{equation}
with the signed derivative $f'(r_h)$, which equals $4\pi sT_+$ at $r_+$ and $-4\pi sT_c$ at $r_c$. Hence $as=b$ is exactly the condition that each horizon obey its own first law, $dE=T_+dS_+$ at the black-hole horizon and $dE=-T_c\,dS_c$ at the cosmological horizon, with the minus sign of Refs.~\cite{GH1977,BJSV2023}. For RN--dS at $(r_+,r_c,Q)=(1,3,0.4)$, for example, $dM/dr_c=-0.4920=-T_c\,dS_c/dr_c$. Hence $F$ is universal, and equal to $F_{\rm cold}$, for every normalization in which the single-horizon first laws hold, and it depends on the normalization otherwise. Three cases illustrate this.

\emph{(a) Bumblebee gravity.} For $ds^2=-f_{\rm b}\,dt^2+(1+L)\,dr^2/f_{\rm b}+r^2d\Omega_2^2$, assumption (A2) fails, since $g_{tt}g_{rr}=-(1+L)$. The surface gravity of $\partial_t$ gives the physical temperature $T_h=|f_{\rm b}'(r_h)|/(4\pi s)$ with $s=\sqrt{1+L}$~\cite{Visser1992,KanziSakalli2019}. The Iyer--Wald analysis of the uncharged, asymptotically flat solution gives the energy $E=sM$ and entropy $S_h=(1+L)\pi r_h^2$~\cite{An2024}, i.e.\ $a=s$ and $b=s^2$, so $as=b$. %REV2 N5
For the charged dS solution we assume the same normalization. This is an assumption, supported by Eq.~\eqref{eq:singlefirstlaw}, which shows that each horizon then obeys its own first law; it is not a derivation of the Iyer--Wald charges for the charged dS solution. The spacetime then behaves as a member of the class: $F=F_{\rm cold}$ and $\Teff=T_c+T_+/x^4$ in terms of the physical temperatures. Compared with $T_h=|f_{\rm b}'|/4\pi$, $E=M$, $S=\mathcal{A}/4$ (for which also $as=b=1$; $\mathcal{A}=4\pi r_h^2$ is the horizon area), every temperature is multiplied by $1/s$ and every entropy and heat capacity by $1+L$, while the positions in $x$ of all zeros, poles and extrema are unchanged.

\emph{(b) Mixed bumblebee normalization.} With the physical temperature but $E=M$ and $S=\mathcal{A}/4$, one has $a=b=1$ and $as=s\neq b$, so the single-horizon first law fails. The cold calibration then gives $G=2s\,x^5/(1-x^3)$, whose solution with $F(0)=1$ is
\begin{equation}
F_L(x)=\Big(1+\tfrac35s\Big)\left(1-x^3\right)^{2/3}+s\left(\frac{2}{5(1-x^3)}-1\right) .
\label{eq:FL}
\end{equation}
$F_L$ reduces to Eq.~\eqref{eq:Fcold} at $L=0$ but otherwise depends on $L$, and the leading-order zero of $C_\Lambda$ (Section~\ref{sec:general}) moves to $x_z^{(0)}=0.80549$, $0.79828$, $0.79198$, $0.77901$ for $L=0.1$, $0.3$, $0.5$, $1$.

\emph{(c) Bousso--Hawking normalization.} In the Bousso--Hawking normalization~\cite{BH1996} the Killing vector is normalized to unit norm at the radius $r_0$ between the horizons where $f'(r_0)=0$, so both temperatures acquire the factor $f(r_0)^{-1/2}$, while the energy is still $M$. Since $\Teff=(\partial M/\partial S)_V$ does not involve the temperatures, Theorem~\ref{thm:structure} still gives $\Teff=\Omega(x)(T_c+T_+/x^4)$ in terms of the $\partial_t$ temperatures, and the cold calibration $\Teff=T_c\,f(r_0)^{-1/2}$ fixes $\Omega=f(r_0)^{-1/2}$ at the cold state. This value depends on $H$ and on $M$, $\Lambda$ and the $\{q_i\}$ at that state, not on $x$ alone: at $x=0.5$, for example, $f(r_0)^{-1/2}=5.9017$ on the cold line of RN--dS and $5.7054$ on that of Conformal Killing gravity with $\lambda r_c^4=0.1$ (Appendix~\ref{app:verification}). The calibration then gives a different $F$ for each metric, and $F$ is no longer universal. Universality holds for the normalization of Eq.~\eqref{eq:temps}, which is the one used throughout the literature on this method.

\section{Structure of $\Teff$ and the calibration}
\label{sec:structure}

\begin{theorem}[Structure of $\Teff$]
\label{thm:structure}
For every spacetime satisfying (A1)--(A4),
\begin{equation}
\Teff=\Omega(x)\left(T_c+\frac{T_+}{x^4}\right),\qquad
\Omega(x)=\frac{2x^5}{(1-x^3)\,G(x)} .
\label{eq:main}
\end{equation}
$\Omega$ depends only on $x$ and on $F$, and hence only on the calibration.
\end{theorem}

\begin{proof}
Rewrite Eq.~\eqref{eq:Teffgeneral} by taking out a factor $x^4$.
\end{proof}

Eq.~\eqref{eq:main} separates $\Teff$ into a kinematic factor fixed by the two horizon temperatures, with a universal weighting $1:x^{-4}$, and a single function $\Omega$ fixed by the calibration.

\subsection{Cold calibration}

\begin{corollary}
\label{cor:cold}
Under the assumptions of Theorem~\ref{thm:universal}, $\Omega=1$ and
\begin{equation}
\Teff=T_c+\frac{T_+}{x^4}
\label{eq:Teffcold}
\end{equation}
for every spacetime in the class.
\end{corollary}

\begin{proof}
At a cold state Eq.~\eqref{eq:main} gives $\Teff=\Omega(x)T_c$, so Eq.~\eqref{eq:coldcond} gives $\Omega(x)=1$. Since $\Omega$ depends only on $x$, the same value holds at every state with that $x$, and a cold state exists for every $x$ by the hypothesis of Theorem~\ref{thm:universal}.
\end{proof}

\subsection{Lukewarm calibration}

Romans~\cite {Romans1992} identified the cold and lukewarm configurations of RN--dS. The lukewarm calibration of Refs.~\cite{ZZM2016,MZC2018,DLZ2022} requires
\begin{equation}
\Teff\big|_{T_+=T_c=T}=T .
\label{eq:lukecond}
\end{equation}

\begin{corollary}
\label{cor:luke}
Suppose that for every $x\in(0,1)$ there is a lukewarm state with $T\neq0$. Then $\Omega=x^4/(1+x^4)$,
\begin{equation}
\Teff=\frac{x^4T_c+T_+}{1+x^4}
\label{eq:Teffluke}
\end{equation}
for every spacetime in the class, and with $F=1+x^2+f_0$, $f_0(0)=0$,
\begin{equation}
f_0(x)=\frac85\left(1-x^3\right)^{2/3}-\frac{2\left(4-5x^3-x^5\right)}{5(1-x^3)} .
\label{eq:f0}
\end{equation}
\end{corollary}

\begin{proof}
At a lukewarm state Eq.~\eqref{eq:main} gives $\Teff=\Omega\,T(1+x^4)/x^4$, so Eq.~\eqref{eq:lukecond} gives $\Omega=x^4/(1+x^4)$, extended to all states as before. Equivalently $G=2x(1+x^4)/(1-x^3)$. Substituting $F=1+x^2+f_0$ gives
\begin{equation}
f_0'+\frac{2x^2}{1-x^3}f_0=\frac{2x^2\left(2x^3+x^2-1\right)}{(1-x^3)^2}.
\end{equation}
With the integrating factor $u^{-2/3}$, the right-hand side integrates term by term. The $x^4$ term uses $\tfrac{d}{dx}\big[x^5u^{-5/3}\big]=5x^4u^{-8/3}$, which follows from $u+x^3=1$. The result is $f_0u^{-2/3}=\tfrac25(1+x^5)u^{-5/3}-2u^{-2/3}+C$, and $f_0(0)=0$ gives $C=8/5$.
\end{proof}

Eq.~\eqref{eq:f0} is the function found for RN--dS in Ref.~\cite{DLZ2022}, its Eq.~(19), and for massive gravity in Ref.~\cite{MZC2018}, its Eqs.~(27)--(28). The published version of Ref.~\cite{ZZM2016}, which introduced the lukewarm calibration, gives a differential equation with a different right-hand side, its Eq.~(14), and a correspondingly different solution, its Eq.~(15). This is not a misprint of Eq.~(14), whose solution is Eq.~(15), but a different result. Its definitions of $S$, $V$, $M$ and $\Teff$ (its Eqs.~(6)--(8)) are those of Section~\ref{sec:class}, and with them the lukewarm condition gives Eq.~\eqref{eq:f0}. Its Eq.~(14) follows instead from the lukewarm temperature written in its Eq.~(13), $(1-x)/[2\pi(1+x^2)r_c]$, whereas the lukewarm states of RN--dS, $M=Q=xr_c/(1+x)$, have $T_+=T_c=(1-x)/[2\pi(1+x)^2r_c]$. The later papers~\cite{MZC2018,DLZ2022} of the same group use Eq.~\eqref{eq:f0}, and so do we.
Ref.~\cite{MZC2018} argued that it takes the same form in both cases. Corollary~\ref{cor:luke} proves this for the whole class.

\subsection{Incompatibility}
\label{sec:conflict}

\begin{figure}[ht]
\centering
\includegraphics[width=0.95\textwidth]{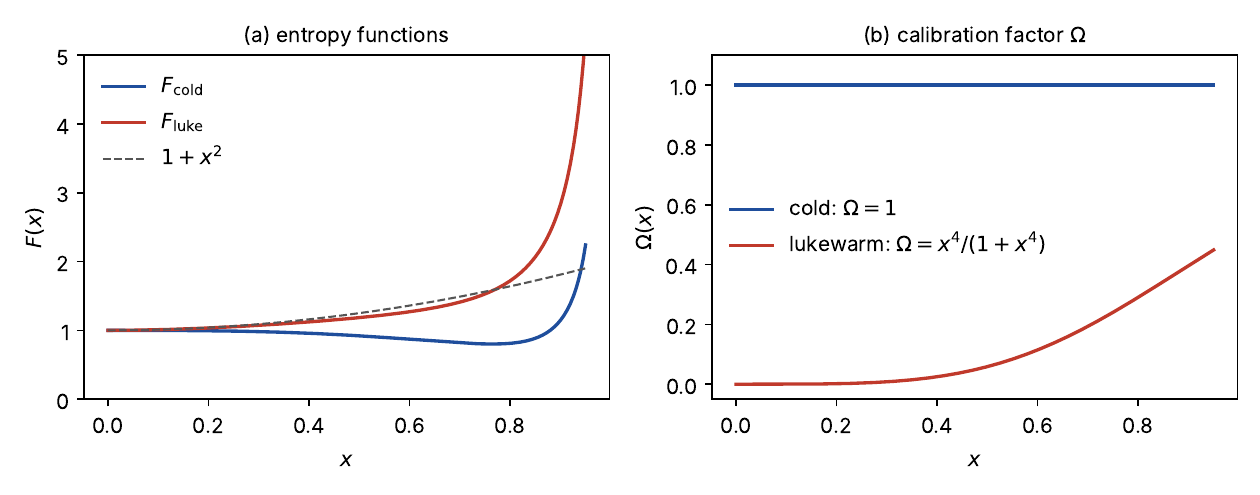}
\caption{(a) Entropy functions $F_{\rm cold}$, Eq.~\eqref{eq:Fcold}, and $F_{\rm luke}=1+x^2+f_0$, Eq.~\eqref{eq:f0}, with the sum of the horizon areas $1+x^2$ for reference. Both are the same for every spacetime in the class. (b) The calibration factor $\Omega(x)$ of Eq.~\eqref{eq:main} for the two calibrations.}
\label{fig:FOmega}
\end{figure}

The two entropy functions differ by
\begin{equation}
F_{\rm luke}-F_{\rm cold}=\frac{x^2\left(5-3x^3\right)}{5\left(1-x^3\right)} ,
\label{eq:DeltaF}
\end{equation}
which is positive on $(0,1)$. The two functions and the corresponding factors $\Omega$ are shown in Fig.~\ref{fig:FOmega}.

\begin{corollary}
\label{cor:conflict}
For any spacetime in the class that admits both cold and lukewarm states, no function $F$ satisfies both Eq.~\eqref{eq:coldcond} and Eq.~\eqref{eq:lukecond}. With the cold calibration, the effective temperature at a lukewarm state is
\begin{equation}
\Teff=T\left(1+\frac{1}{x^4}\right)\neq T .
\label{eq:conflict}
\end{equation}
\end{corollary}

\begin{proof}
The two calibrations require $\Omega=1$ and $\Omega=x^4/(1+x^4)$, which differ for every $x\in(0,1)$. Eq.~\eqref{eq:conflict} follows from Eq.~\eqref{eq:Teffcold} with $T_+=T_c=T$.
\end{proof}

\begin{remark}
Eq.~\eqref{eq:conflict} agrees with Eq.~(26) of Ref.~\cite{Zhen2025}, where it was obtained for RN--dS with the cold calibration and interpreted as an interaction between the horizons. Corollary~\ref{cor:conflict} shows that the same result also expresses a conflict with the lukewarm calibration of Refs.~\cite{ZZM2016,MZC2018,DLZ2022}.
\end{remark}

\section{General results}
\label{sec:general}

\subsection{Smarr relation}

\begin{remark}[Smarr relation]
\label{lem:smarr}
Let each parameter $q_i$ have length dimension $d_i$, so that $f$ is invariant under $r\to\lambda r$, $M\to\lambda M$, $\Lambda\to\lambda^{-2}\Lambda$, $q_i\to\lambda^{d_i}q_i$. Then
\begin{equation}
M=2\Teff S-3\Peff V+\sum_i d_i\,q_i\varphi_i .
\label{eq:smarr}
\end{equation}
Since $F$ depends only on the scale-invariant ratio $x$, $S$ and $V$ have weights $2$ and $3$, and $M(S,V,\{q_i\})$ is homogeneous of degree $1$; Euler's theorem and Eq.~\eqref{eq:Teffdef} give Eq.~\eqref{eq:smarr}.
\end{remark}

The relation follows from scaling and is not a new result~\cite{Smarr1973,KRT2009,KMT2017}. For RN--dS ($d_Q=1$) and Conformal Killing gravity ($d_Q=1$, $d_\lambda=-4$), it reproduces $M=2\Teff S-3\Peff V+\varphi_QQ-4\lambda\varphi_\lambda$, which is Eq.~(40) of Ref.~\cite{Paper2}, where $\varphi_\lambda$ is denoted $\Psi_\lambda$.
The residuals are listed in Appendix~\ref{app:verification}.

\subsection{Heat capacity}

The heat capacity is
\begin{equation}
C_X=\Teff\left(\frac{\partial S}{\partial\Teff}\right)_X=\Teff\,\frac{(dS/dx)_X}{(d\Teff/dx)_X} ,
\label{eq:CX}
\end{equation}
where $X$ is held fixed along the path together with $\{q_i\}$. We consider $X=V$, consistent with Eq.~\eqref{eq:Teffdef}, and $X=\Lambda$, as used in Refs.~\cite{Zhen2025,Paper1,Paper2,Paper3}.

\paragraph{Fixed volume.}
\begin{lemma}
\label{lem:dSV}
At fixed $V$, $r_c=R\,u^{-1/3}$ with $R\equiv(3V/4\pi)^{1/3}$, and $(dS/dx)_V=\pi R^2G(x)\,u^{-5/3}$.
\end{lemma}

\begin{proof}
$S=\pi R^2Fu^{-2/3}$, so $dS/dx=\pi R^2u^{-5/3}\left[uF'+2x^2F\right]$.
\end{proof}

For the two calibrations,
\begin{equation}
\left(\frac{dS}{dx}\right)_V^{\rm cold}=\frac{2\pi R^2x^5}{u^{8/3}},\quad
\left(\frac{dS}{dx}\right)_V^{\rm luke}=\frac{2\pi R^2x\left(1+x^4\right)}{u^{8/3}},
\label{eq:dSVcal}
\end{equation}
both the same for every spacetime in the class and both positive on $(0,1)$.

\begin{remark}
\label{cor:CV}
For every spacetime in the class and either calibration, $C_V$ has no zeros in the physical domain $T_+\ge0$, because there $T_c>0$ and $T_+\ge0$, so $\Teff>0$ by Eqs.~\eqref{eq:Teffcold}--\eqref{eq:Teffluke}, and $(dS/dx)_V>0$ by Eq.~\eqref{eq:dSVcal}. It changes sign only through poles, located where $(d\Teff/dx)_V=0$.
\end{remark}

\paragraph{Fixed cosmological constant.} 
The heat capacity at fixed $\Lambda$ is meaningful with $\Teff$ defined at fixed $V$. Since $M$ is a function of $(S,V,\{q_i\})$, the first law~\eqref{eq:firstlaw} holds for every variation at fixed $\{q_i\}$, not only at fixed $V$. Along a path at fixed $\Lambda$ and $\{q_i\}$ both $S$ and $V$ change, $dM=\Teff\,dS-\Peff\,dV$, and the heat absorbed is $\Teff\,dS$. Hence $C_\Lambda=\Teff(\partial S/\partial\Teff)_{\Lambda,\{q\}}$ is the heat capacity of the effective system along that path, with $\Teff$ still given by Eq.~\eqref{eq:Teffdef}, in the same way as $C_P$ is defined with the temperature $(\partial U/\partial S)_V$ of an ordinary system.

\begin{figure}[ht]
\centering
\includegraphics[width=0.6\textwidth]{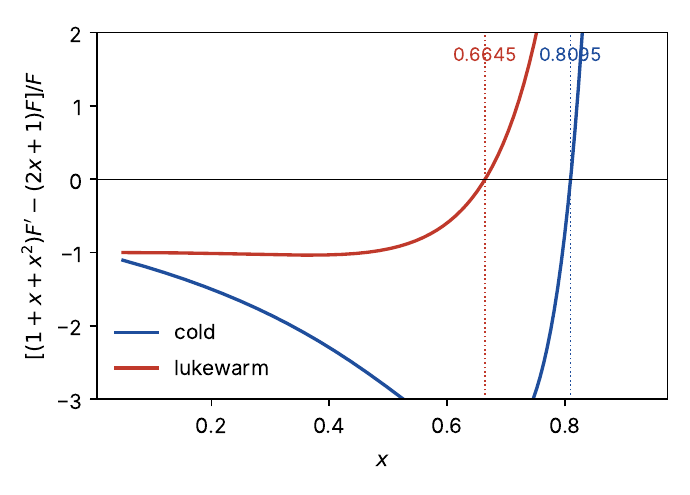}
\caption{The universal zero condition~\eqref{eq:xz}, plotted as $[(1+x+x^2)F'-(2x+1)F]/F$ for the cold and lukewarm entropy functions. The roots are the zeros of $C_\Lambda$ when $\chi$ is constant along the path, Eq.~\eqref{eq:xzvalues}, for every spacetime in the class.}
\label{fig:zero}
\end{figure}

Along the path, Eq.~\eqref{eq:rcLambda} can be written exactly as
\begin{equation}
r_c^2=\frac{3k\left[1+\chi(x)\right]}{\Lambda A},\qquad
\chi(x)\equiv\frac{r_cH(r_c)-r_+H(r_+)}{k\,r_c(1-x)},
\label{eq:chi}
\end{equation}
with $k>0$ and $r_c=r_c(x)$ on the right. Since $dS/dx=\pi\left(2r_cr_c'F+r_c^2F'\right)$ and $2r_c'/r_c=\chi'/(1+\chi)-A'/A$, $C_\Lambda$ vanishes exactly where
\begin{equation}
\left(1+x+x^2\right)F'-(2x+1)F=-A\,F\,\frac{\chi'}{1+\chi} .
\label{eq:xzexact}
\end{equation}
The right-hand side vanishes when $\chi$ is constant along the path, for example, for $H=0$ or for a constant shift of $k$. The zero condition then becomes
\begin{equation}
\left(1+x+x^2\right)F'=(2x+1)F ,
\label{eq:xz}
\end{equation}
which contains no parameter of the spacetime. Its root in $(0,1)$ depends only on the calibration:
\begin{equation}
x_z^{(0)}=0.80953\ \ \text{(cold)},\qquad x_z^{(0)}=0.66450\ \ \text{(lukewarm)} .
\label{eq:xzvalues}
\end{equation}
The two roots are shown in Fig.~\ref{fig:zero}.

The size of the correction is controlled by $\chi'/(1+\chi)$, not by $\chi$ itself. Table~\ref{tab:chi} lists both for the examples of Section~\ref{sec:examples}. Over the stable windows $|\chi|$ lies between $0.08$ and $0.53$, so $\chi$ is not small. At the zero, however, $|\chi'/(1+\chi)|$ is only $1.6$--$6.5\%$ of the parameter-free term $A'/A=(2x+1)/A$, and $x_z$ lies within $0.0021$ ($0.26\%$) of $x_z^{(0)}$ in these ten configurations. 

The deviation grows with the charge. For RN--dS at $\Lambda=0.3$ it is $-0.43\%$ at $Q=0.815$ and $-0.76\%$ at $Q=0.86$ (last row of Table~\ref{tab:chi}), where the stable window has shrunk to $0.673<x<0.803$. Eq.~\eqref{eq:xz} was obtained for the nonlinearly charged dS black hole in Ref.~\cite{Paper3}; Eq.~\eqref{eq:xzexact} extends it to the whole class and shows when it is exact.

\begin{table}[ht]
\centering
\small
\renewcommand{\arraystretch}{1.15}
\begin{tabular}{lccc}
\toprule
Configuration & $|\chi|$ on $(x_f,x_z)$ & $\dfrac{|\chi'/(1+\chi)|}{A'/A}\Big|_{x_z}$ & $x_z-x_z^{(0)}$\\
\midrule
RN--dS, $Q=0.5$  & $0.083$--$0.161$ & $0.016$ & $-0.0005$\\
RN--dS, $Q=0.75$ & $0.220$--$0.283$ & $0.065$ & $-0.0021$\\
CKG, $\mu=1.6$   & $0.138$--$0.220$ & $0.019$ & $-0.0006$\\
CKG, $\mu=5$     & $0.290$--$0.390$ & $0.036$ & $-0.0011$\\
CKG, $\mu=12$    & $0.436$--$0.534$ & $0.049$ & $-0.0016$\\
NLE, $(0.25,1.2)$ & $0.155$--$0.230$ & $0.037$ & $-0.0012$\\
NLE, $(0.35,1.2)$ & $0.141$--$0.218$ & $0.032$ & $-0.0010$\\
NLE, $(0.45,1.2)$ & $0.135$--$0.213$ & $0.030$ & $-0.0010$\\
NLE, $(0.35,1.6)$ & $0.111$--$0.190$ & $0.023$ & $-0.0007$\\
NLE, $(0.35,2.0)$ & $0.093$--$0.172$ & $0.019$ & $-0.0006$\\
\midrule
RN--dS, $Q=0.86$ & $0.343$--$0.372$ & $0.185$ & $-0.0062$\\
\bottomrule
\end{tabular}
\caption{Size of the deviation from the parameter-free zero condition, cold calibration. $\chi$ from Eq.~\eqref{eq:chi}, evaluated on the stable window $x_f<x<x_z$ of $C_\Lambda$. RN--dS: $\Lambda=0.3$. CKG: parameters of Ref.~\cite{Paper2} (Section~\ref{sec:examples}). NLE: $(\alpha,\omega)$ of Ref.~\cite{Paper3}, $q=0.75$.}
\label{tab:chi}
\end{table}

\paragraph{Positive windows and maxima of $C_\Lambda$.} We call a branch stable where $C_\Lambda>0$, for $x$ above the cold boundary $x_A$. With the cold calibration, there is one such window, $x_f<x<x_z$, between the pole and the zero of $C_\Lambda$; between $x_A$ and $x_f$ the heat capacity is negative. With the lukewarm calibration, there are two: a narrow window $x_A<x<x_1$, between the cold boundary and a zero $x_1$ of $dS/dx$ below the pole, and the main window $x_f<x<x_z$. At $x_A$, the lukewarm temperature is $\Teff=x_A^4T_c/(1+x_A^4)>0$ and $\Teff$ increases with $x$, so the narrow window is present in every lukewarm case we examined. In the narrow window, both $S$ and $\Teff$ increase with $x$. At $x_1$ the effective entropy reaches a local maximum along the path and then decreases up to the pole, where $C_\Lambda<0$; the temperatures in the window are small ($\Teff<0.025$ for RN--dS at $Q=0.5$, against $0.205$ at the cold maximum). The narrow-window maximum is therefore a feature of the lukewarm entropy function close to the cold boundary, and we report it without interpreting it as a two-level Schottky peak. Table~\ref{tab:peaks} lists both windows and their maxima. With the cold calibration, RN--dS at $\Lambda=0.3$ has a maximum at $Q=0.5$ but none at $Q=0.75$. With the lukewarm calibration, every example has a maximum in the narrow window, while the main window has one for Conformal Killing gravity but none for RN--dS at $\Lambda=0.3$. The existence and the position of a Schottky-like peak therefore depend on the parameters and on the calibration.

\subsection{Limits}
\label{sec:limits}

\paragraph{Small black hole, $x\to0$.}
\begin{align}
F_{\rm cold}&=1-\tfrac23x^3+\tfrac29x^6+O(x^9), \label{eq:Fcold0}\\
F_{\rm luke}&=1+x^2-\tfrac23x^3+\tfrac25x^5+\tfrac29x^6+O(x^8). \label{eq:Fluke0}
\end{align}
The lukewarm function gives $S\simeq\pi r_c^2+\pi r_+^2$ for a small black hole, the sum of the two horizon entropies. The cold function has no $x^2$ term, so the black-hole area does not appear at leading order. By Eq.~\eqref{eq:DeltaF}, this difference is exactly the leading term $x^2$ of $F_{\rm luke}-F_{\rm cold}$.

In both cases $S\to\pi r_c^2$ as $x\to0$, but $\Teff$ does not tend to $T_c$. For a small black hole at fixed $r_c$ with $H$ negligible near $r_+$, $T_+\simeq1/(4\pi r_+)=1/(4\pi x\,r_c)$, while $T_c$ tends to the de~Sitter value $1/(2\pi r_c)$. Eq.~\eqref{eq:Teffcold} then gives $\Teff\simeq x^{-5}/(4\pi r_c)$, which diverges, and Eq.~\eqref{eq:Teffluke} gives $\Teff\to T_+$. Neither calibration recovers the pure de~Sitter temperature in this limit. A second property that bears on the choice is subadditivity of the entropy. The effective entropy is smaller than the sum of the horizon entropies, $S<\pi r_+^2+\pi r_c^2$, i.e.\ $F<1+x^2$, for $x<0.9411$ with the cold calibration and for $x<0.7696$ (where $f_0<0$) with the lukewarm one; beyond these values it is larger. We record these properties as facts relevant to the choice of calibration, without claiming that they decide it. For comparison, the on-shell Euclidean action of RN--dS is minus the sum of the two horizon entropies for arbitrary mass and charge~\cite{MvdSV2024}, which corresponds to $F=1+x^2$; neither $F_{\rm cold}$ nor $F_{\rm luke}$ has this form.

\paragraph{Nariai limit, $x\to1$.} Both functions have a universal simple pole,
\begin{equation}
F\simeq\frac{c}{1-x},\qquad c=\frac{2}{15}\ \text{(cold)},\qquad c=\frac{4}{15}\ \text{(lukewarm)} ,
\label{eq:Nariai}
\end{equation}
while $V\propto r_c^3(1-x^3)$ vanishes linearly at fixed $r_c$. At fixed $\Lambda$, both horizon radii remain finite, so the divergence of $S$ comes from $F$ and not from the horizon areas. At fixed $V$, $r_c=Ru^{-1/3}$ diverges as well and $S\propto u^{-5/3}$.

The heat capacity at fixed $\Lambda$ has a universal leading behaviour near the Nariai limit. Let $r_N$ be the common radius at $x=1$, the Nariai limit~\cite{Nariai1950,Nariai1951,GP1983,BH1996}, and assume $f''(r_N)\neq0$, which excludes the ultracold point.

\begin{proposition}
\label{prop:Nariai}
As $x\to1$ at fixed $\Lambda$ and $\{q_i\}$,
\begin{equation}
(1-x)\,C_\Lambda\to-\pi r_N^2\,c ,
\label{eq:CNariaix}
\end{equation}
with $c$ from Eq.~\eqref{eq:Nariai}, and
\begin{equation}
C_\Lambda\simeq-\frac{r_N^3\,|f''(r_N)|}{30\,\Teff}
\label{eq:CNariaiT}
\end{equation}
for both calibrations.
\end{proposition}

\begin{proof}
Near the double root, $f(r)\simeq\frac12f''(r_N)(r-r_+)(r-r_c)$, so $T_+\simeq T_c\simeq|f''(r_N)|\,(r_c-r_+)/(8\pi)$ and $T_c+T_+\simeq a\,(1-x)$ with $a=r_N|f''(r_N)|/(4\pi)$. Since $\Omega_{\rm cold}(1)=1$ and $\Omega_{\rm luke}(1)=1/2$, Eqs.~\eqref{eq:Teffcold}--\eqref{eq:Teffluke} give $\Teff\simeq a(1-x)$ and $\Teff\simeq\frac a2(1-x)$ respectively. Since $r_c\to r_N$, Eq.~\eqref{eq:Nariai} gives $dS/dx\simeq\pi r_N^2c/(1-x)^2$. Inserting both into Eq.~\eqref{eq:CX} gives Eq.~\eqref{eq:CNariaix}, in which the slope of $\Teff$ cancels. Eliminating $1-x$ in favour of $\Teff$ gives $C_\Lambda\simeq-\pi r_N^2\,c\,\sigma/\Teff$, where $\sigma$ is the slope of $\Teff$. For the cold calibration $c\,\sigma=\frac{2}{15}a$; for the lukewarm calibration $c\,\sigma=\frac{4}{15}\cdot\frac a2=\frac{2}{15}a$. Both give Eq.~\eqref{eq:CNariaiT}.
\end{proof}

Two conclusions follow. First, the divergence is not an artifact of the variable $x$. $C=\Teff\,dS/d\Teff$ does not depend on how the path is parametrized, and Eq.~\eqref{eq:CNariaiT} shows that it diverges as $1/\Teff$. Its origin is the pole of $F$: if $F$ were finite at $x=1$, $C_\Lambda$ would vanish as $\Teff\to0$. Second, the strength of the divergence is the same for both calibrations. The doubled pole of $F_{\rm luke}$ is compensated exactly by the halved slope of its $\Teff$, so the Nariai limit is the one regime where the calibration does not matter at leading order. For RN--dS with $Q=0.5$, $\Lambda=0.3$ one has $r_N^2=3.0611$ and $f''(r_N)=-0.54664$, and Eqs.~\eqref{eq:CNariaix}--\eqref{eq:CNariaiT} give $-1.2822$ (cold), $-2.5645$ (lukewarm) and $C_\Lambda\Teff\to-0.097588$ (both), in agreement with direct computation (Appendix~\ref{app:verification}).

\section{Worked examples}
\label{sec:examples}

We apply the general results to four spacetimes. Table~\ref{tab:examples} shows how they fit the class. For each, $\Teff$ was computed directly from Eq.~\eqref{eq:Teffdef}, by solving $f(r_+)=f(r_c)=0$ for $M$ and $\Lambda$ and differentiating along the fixed-volume path, and compared with Eq.~\eqref{eq:Teffcold}. The difference simplifies symbolically to zero in every case (Appendix~\ref{app:verification}).

\begin{table}[ht]
\centering
\small
\renewcommand{\arraystretch}{1.4}
\setlength{\tabcolsep}{5pt}
\begin{tabular}{llll}
\toprule
Spacetime & $k$ & $H(r)$ & $\Teff-(T_c+T_+/x^4)$ \\
\midrule
RN--dS
& $1$
& $Q^2/r^2$
& $0$ \\

\parbox[t]{3.2cm}{Conformal Killing gravity~\cite{Paper2}}
& $1$
& $q^2/r^2-\lambda r^4/5$
& $0$ \\

\parbox[t]{3.2cm}{Nonlinear electrodynamics~\cite{Paper3}}
& $1+2\alpha q$
& \parbox[t]{5cm}{$q^2/r^2$ \\ ($\Lambda\to\Lambda_{\rm eff}=\Lambda+\alpha^2$)}
& $0$ \\

\parbox[t]{3.2cm}{Bumblebee gravity~\cite{Paper1}}
& $1$
& \parbox[t]{5cm}{$Q_{\rm eff}^2/r^2$, with \\ $\Lambda\to\Lambda_{\rm eff}=(1+L)\Lambda$}
& \parbox[t]{3cm}{$0$ in normalization (ii)$^{\dagger}$ (Section~\ref{sec:bumblebee})} \\

\bottomrule
\end{tabular}

\caption{Spacetimes of the class~\eqref{eq:metric} for which
Corollary~\ref{cor:cold} was verified directly. The metric functions
are Eq.~(16) of Ref.~\cite{Paper2}, Eq.~(2) of Ref.~\cite{Paper3},
and Eq.~\eqref{eq:bumblebee}. For nonlinear electrodynamics, the term
$-\alpha^2r^2/3$ combines with the cosmological term
$-r^2/\ell^2$ into $-\Lambda_{\rm eff}r^2/3$, which does not depend
on $M$, so assumption (A1) holds with $\Lambda_{\rm eff}$ in place
of $\Lambda$. For bumblebee gravity
$Q_{\rm eff}^2=2(1+L)Q^2/(2+L)$.
$^{\dagger}$In the consistent normalization (ii) of
Section~\ref{sec:bumblebee}, with the physical temperatures of
Section~\ref{sec:normalization}.}
\label{tab:examples}
\end{table}

\subsection{RN--dS}
With $k=1$ and $H=Q^2/r^2$, Eqs.~\eqref{eq:Mrcx}--\eqref{eq:Lrcx} reduce to the standard RN--dS expressions, and Theorem~\ref{thm:universal} and Corollary~\ref{cor:luke} reproduce the cold function of Ref.~\cite{Zhen2025} and the lukewarm function of Refs.~\cite{ZZM2016,DLZ2022}. Both identities~\eqref{eq:Teffcold} and~\eqref{eq:Teffluke} hold exactly. RN--dS is used as the example for the calibration comparison in Section~\ref{sec:calibration}.

\subsection{Conformal Killing gravity}\label{sec:CKG}
Conformal Killing gravity~\cite{Harada2023,MM2023,MM2024} admits charged static spherically symmetric solutions~\cite{CN2024,Barnes2023}. With $k=1$ and $H=q^2/r^2-\lambda r^4/5$, Eq.~\eqref{eq:rcLambda} becomes the cubic $\mu BY^3+5AY^2-5Y+5\tilde q^2/x=0$ for $Y=r_c^2/\ell^2$ used in Ref.~\cite{Paper2}, with $B=1+x+x^2+x^3+x^4$, $\mu=\lambda\ell^4$ and $\tilde q=q/\ell$. Corollary~\ref{cor:cold} reproduces its effective temperature, Eq.~(31) of Ref.~\cite{Paper2}, and Theorem~\ref{thm:universal} its entropy function, Eq.~(32). The Smarr relation of Remark~\ref{lem:smarr} is Eq.~(40). 

Table~\ref{tab:peaks} gives the canonical heat capacity $C_\Lambda$ obtained from the general formulas for the parameters of Ref.~\cite{Paper2}: $q=0.06$, $\ell^2=12\eta q^2$ with $\eta=70$, and $\mu=1.6$, $5$, $12$, along the path at fixed $q$, $\ell$ and $\lambda$ used there (its Eq.~(44)), with the physical (largest) root of the cubic. The maximum lies at $x_{\rm peak}=0.595$, $0.604$, $0.609$, with $\Teff=0.496$, $0.550$, $0.635$ and $C_{\max}=0.632$, $0.511$, $0.400$; $C_\Lambda$ vanishes at $x_z\simeq0.808$ and is negative beyond it, as Eq.~\eqref{eq:xzexact} and Proposition~\ref{prop:Nariai} require. Table~I of Ref.~\cite{Paper2} lists instead $x_{\rm peak}=0.749$, $0.749$, $0.748$, $\Teff=0.240$, $0.294$, $0.349$ and $C_{\max}=0.114$, $0.0765$, $0.0537$. These values could not be reproduced. Along the fixed-$\Lambda$ path, the maximum lies at $x=0.595$--$0.609$, also for $\eta$ between $30$ and $140$; on the other positive roots of the cubic and along fixed-volume paths there is no interior maximum at all (Appendix~\ref{app:verification}). Since the equations in Ref.~\cite{Paper2} agree with ours term by term, the values in Table~\ref{tab:peaks} replace those in Table~I of ~\cite{Paper2}. The qualitative result of Ref.~\cite{Paper2} is unchanged: a single maximum whose height decreases and whose temperature increases with $\mu$.

\subsection{Nonlinear electrodynamics}
For the nonlinearly charged dS black hole~\cite{Akhtar2023} studied in Refs.~\cite{Paper3,Zhen2024CQG}, with $k=1+2\alpha q$, $H=q^2/r^2$ and $\Lambda_{\rm eff}=3/\ell^2+\alpha^2$, Eq.~\eqref{eq:rcLambda} is a quadratic in $r_c^2$. Using only the general results of Sections~\ref{sec:theorem}--\ref{sec:general}, namely $\Teff=T_c+T_+/x^4$, the universal $F_{\rm cold}$ and the path~\eqref{eq:rcLambda} at fixed $q$, $\alpha$ and $\ell$, we recomputed the full canonical heat-capacity structure of Table~IV of Ref.~\cite{Paper3} for all five configurations, with $q=0.75$ and $\ell^2=12\omega q^2$. All $35$ values agree with the published ones to within a relative difference of $4\times10^{-5}$ (Table~\ref{tab:published}). 
Two entries, $\Teff$ at the peak for the first and fifth configurations, recompute as $0.23750$ and $0.26312$ against $0.23749$ and $0.26313$ in Ref.~\cite{Paper3}; Table~\ref{tab:published} gives the recomputed values, rounded consistently. The zeros $x_z$ lie $0.07$--$0.15\%$ below $x_z^{(0)}=0.80953$, as Eq.~\eqref{eq:xzexact} and Table~\ref{tab:chi} explain.

\subsection{Bumblebee gravity}
\label{sec:bumblebee}
Bumblebee gravity~\cite{Casana2018} admits a charged dS black hole~\cite{Liu2025}. The solution studied in Ref.~\cite{Paper1} has
\begin{equation}
ds^2=-f_{\rm b}\,dt^2+\frac{1+L}{f_{\rm b}}\,dr^2+r^2d\Omega_2^2,
\label{eq:bumblebee}
\end{equation}
with $f_{\rm b}(r)=1-2M/r+Q_{\rm eff}^2/r^2-(1+L)\Lambda r^2/3$ and $Q_{\rm eff}^2=2(1+L)Q^2/(2+L)$. (Ref.~\cite{Paper1} denotes this function by $A$.) Thus $f_{\rm b}$ is of the form~\eqref{eq:metric} with $k=1$ and $\Lambda_{\rm eff}=(1+L)\Lambda$, but $g_{tt}g_{rr}=-(1+L)$.

\emph{Which temperature is physical.} The surface gravity of $\partial_t$ gives $T_h=|f_{\rm b}'(r_h)|/(4\pi\sqrt{1+L})$~\cite{Visser1992,KanziSakalli2019}, and this is the physical temperature. Ref.~\cite{Paper1} used $T_h=|f_{\rm b}'(r_h)|/4\pi$ in its Eqs.~(10), (12), (15) and (16), together with $E=M$ and $S=\mathcal{A}/4$, where $\mathcal{A}=4\pi r_h^2$ is the horizon area. Combined with the physical temperature, Section~\ref{sec:normalization} allows two choices of energy and entropy: the consistent one, $E=\sqrt{1+L}\,M$ and $S=(1+L)\,\mathcal{A}/4$, in which each horizon obeys its own first law, and the mixed one, $E=M$ and $S=\mathcal{A}/4$, in which it does not. Table~\ref{tab:paper1} compares the three cases for the parameters of Ref.~\cite{Paper1}: $Q=1$, $L=0.4$ and $\ell^2=12\zeta Q^2/(1+L)$ with $\zeta=30$, $70$, $140$, so that $\Lambda_{\rm eff}=3(1+L)/\ell^2$.

We adopt the consistent normalization, case (ii), in which the bumblebee black hole is a member of the class. As stated in Section~\ref{sec:normalization}, its use for the charged dS solution is an assumption supported by Eq.~\eqref{eq:singlefirstlaw}, not a derivation of the Iyer--Wald charges. Compared with Ref.~\cite{Paper1}, case (i), the positions in $x$ of the cold boundary, the pole, the maximum, and the zero of $C_\Lambda$ are unchanged, and $x_z=0.8094$--$0.8095$ remains at the universal value. Two scales change:
\begin{itemize}
\item every temperature is multiplied by $1/\sqrt{1+L}=0.84515$: the horizon and effective temperatures of Eqs.~(10), (12), (15) and (16) of Ref.~\cite{Paper1}, its Figs.~4--7, and the temperature axis of its Fig.~10;
\item the entropy and every heat capacity are multiplied by $1+L=1.4$: Eqs.~(31), (32) and (36) and Figs.~9 and~10 of Ref.~\cite{Paper1}, and with them the estimate of $N$ in its Eq.~(37).
\end{itemize}
The normalized two-level heat capacity of Fig.~11 of Ref.~\cite{Paper1} and the relation $T_{\rm peak}\simeq0.4167\,\Delta$ of its Eq.~(38) depend only on the ratio of the gap to $\Teff$, and are unchanged. Independently of the normalization, the coefficient $2.7$ in Eqs.~(32), (36) and (37) of Ref.~\cite{Paper1}, $C_{\max}\simeq2.7\,\zeta Q^2/(1+L)$, is not reproduced: the general formulas give $C_{\max}/\zeta=1.49$--$1.52$ in case (i) and $2.09$--$2.13$ in case (ii), against $2.7/(1+L)=1.93$. In the mixed case (iii), $F=F_L$ of Eq.~\eqref{eq:FL} and the zero moves to $x_z=0.795$.

\begin{table}[ht]
\centering
\small
\setlength{\tabcolsep}{3.2pt}
\renewcommand{\arraystretch}{1.2}
\begin{tabular}{cc|cccccc}
\toprule
$\zeta$ & case & $x_A$ & $x_f$ & $x_{\rm peak}$ & $\Teff^{\rm peak}$ & $C_{\max}$ & $x_z$ \\
\midrule
30  & (i)   & 0.08815 & 0.10408 & 0.58333 & 0.05855 & 44.816  & 0.80943 \\
    & (ii)  & 0.08815 & 0.10408 & 0.58333 & 0.04949 & 62.742  & 0.80943 \\
    & (iii) & 0.08815 & 0.10408 & 0.56915 & 0.05703 & 43.926  & 0.79490 \\
\midrule
70  & (i)   & 0.05545 & 0.06554 & 0.58238 & 0.03969 & 105.835 & 0.80949 \\
    & (ii)  & 0.05545 & 0.06554 & 0.58238 & 0.03354 & 148.169 & 0.80949 \\
    & (iii) & 0.05545 & 0.06554 & 0.56812 & 0.03872 & 103.768 & 0.79498 \\
\midrule
140 & (i)   & 0.03844 & 0.04546 & 0.58202 & 0.02842 & 212.611 & 0.80951 \\
    & (ii)  & 0.03844 & 0.04546 & 0.58202 & 0.02402 & 297.656 & 0.80951 \\
    & (iii) & 0.03844 & 0.04546 & 0.56773 & 0.02775 & 208.482 & 0.79500 \\
\bottomrule
\end{tabular}
\caption{Canonical heat capacity of the bumblebee black hole for the parameters of Ref.~\cite{Paper1} ($Q=1$, $L=0.4$), cold calibration. (i)~$T=|f_{\rm b}'|/4\pi$, $E=M$, $S=\mathcal{A}/4$, as in Ref.~\cite{Paper1}; (ii)~physical $T=|f_{\rm b}'|/(4\pi\sqrt{1+L})$ with $E=\sqrt{1+L}\,M$, $S=(1+L)\mathcal{A}/4$, adopted here; (iii)~physical $T$ with $E=M$, $S=\mathcal{A}/4$ and $F=F_L$. $x_A$: cold boundary; $x_f$: pole; $x_z$: zero.}
\label{tab:paper1}
\end{table}

%Table~\ref{tab:peaks} compares the two cases for $Q=0.5$, $\Lambda=0.3$. In case~(a) the zero stays at $0.809$; in case~(b) it moves to $0.805$ ($L=0.1$) and $0.797$ ($L=0.3$), close to the leading-order values above. The bumblebee spacetime is therefore the example of Section~\ref{sec:boundary} for assumption~(A2), provided the surface-gravity temperature is used. 

\begin{table}[htbp]
\centering
\small
\renewcommand{\arraystretch}{1.2}

\setlength{\tabcolsep}{4pt}
\begin{tabular}{l|cccc|ccc|cc}
\toprule
 & & & & & \multicolumn{3}{c|}{maximum on $x_f<x<x_z$} & \multicolumn{2}{c}{maximum on $x_A<x<x_1$}\\
Spacetime and calibration & $x_A$ & $x_1$ & $x_f$ & $x_z$ & $x_{\rm peak}$ & $\Teff$ & $C_{\max}$ & $x_{\rm peak}$ & $C_{\max}$ \\
\midrule
CKG, $\mu=1.6$, cold & 0.04006 & -- & 0.04738 & 0.80892 & 0.59498 & 0.49552 & 0.63163 & -- & -- \\
CKG, $\mu=5$, cold   & 0.04548 & -- & 0.05379 & 0.80839 & 0.60354 & 0.54996 & 0.51077 & -- & -- \\
CKG, $\mu=12$, cold  & 0.05223 & -- & 0.06176 & 0.80797 & 0.60890 & 0.63482 & 0.40045 & -- & -- \\
CKG, $\mu=1.6$, luke & 0.04006 & 0.04190 & 0.06929 & 0.65690 & 0.32268 & 0.14626 & 0.93684 & 0.04099 & $2.56\times10^{-4}$ \\
CKG, $\mu=5$, luke   & 0.04548 & 0.04824 & 0.07866 & 0.64969 & 0.34638 & 0.15488 & 0.66039 & 0.04688 & $3.39\times10^{-4}$ \\
CKG, $\mu=12$, luke  & 0.05223 & 0.05633 & 0.09034 & 0.64349 & 0.36326 & 0.17088 & 0.46626 & 0.05432 & $4.38\times10^{-4}$ \\
\midrule
RN--dS, $Q=0.5$, cold  & 0.20435 & -- & 0.23969 & 0.80902 & 0.58733 & 0.20542 & 2.28325 & -- & -- \\
RN--dS, $Q=0.5$, luke  & 0.20435 & 0.23178 & 0.31970 & 0.65729 & none & -- & -- & 0.21844 & 0.04859 \\
RN--dS, $Q=0.75$, cold & 0.40687 & -- & 0.46787 & 0.80746 & none & -- & -- & -- & -- \\
RN--dS, $Q=0.75$, luke & 0.40687 & 0.48352 & 0.55056 & 0.62056 & none & -- & -- & 0.44545 & 0.15438 \\
\bottomrule
\end{tabular}
\caption{Canonical heat capacity $C_\Lambda$ from the general formulas, with both positive windows. CKG: $q=0.06$, $\ell^2=12\eta q^2$, $\eta=70$, parameters of Ref.~\cite{Paper2}; the cold rows replace Table~I of Ref.~\cite{Paper2}. RN--dS: $\Lambda=0.3$. $x_A$: cold boundary ($T_+=0$); $x_1$: zero of $C_\Lambda$ below the pole (lukewarm calibration only); $x_f$: pole; $x_z$: upper zero. $C_\Lambda>0$ on $x_f<x<x_z$ for both calibrations and, for the lukewarm calibration, also on $x_A<x<x_1$. ``none'': no interior maximum on that window. The bumblebee black hole is in Tables~\ref{tab:paper1} and~\ref{tab:external}.}
\label{tab:peaks}
\end{table}

\begin{table}[ht]
\centering
\small
\renewcommand{\arraystretch}{1.2}
\begin{tabular}{cc|ccccccc}
\toprule
$\alpha$ & $\omega$ & $x_A$ & $x_f$ & $\Teff^{\max}$ & $x_{\rm peak}$ & $T_{\rm eff,peak}$ & $C_{\max}$ & $x_z$ \\
\midrule
0.25 & 1.2 & 0.31115 & 0.36166 & 1.3180 & 0.58205 & 0.23750 & 2.04136 & 0.80836 \\
0.35 & 1.2 & 0.29098 & 0.33887 & 2.1206 & 0.58529 & 0.27393 & 2.00804 & 0.80851 \\
0.45 & 1.2 & 0.28204 & 0.32873 & 2.8295 & 0.58615 & 0.31420 & 1.90691 & 0.80857 \\
0.35 & 1.6 & 0.24620 & 0.28784 & 4.6273 & 0.58765 & 0.26872 & 2.53951 & 0.80880 \\
0.35 & 2.0 & 0.22033 & 0.25812 & 7.6688 & 0.58761 & 0.26312 & 2.99871 & 0.80894 \\
\bottomrule
\end{tabular}
\caption{Canonical heat-capacity structure of the nonlinearly charged dS black hole ($q=0.75$), recomputed from the general formulas of this paper. All $35$ values agree with Table~IV of Ref.~\cite{Paper3} to a relative difference of at most $4\times10^{-5}$; the two entries of $T_{\rm eff,peak}$ that differ in the last digit (rows 1 and 5) are given as recomputed. $x_A$: cold boundary; $x_f$: pole of $C$, where $\Teff$ has its local maximum $\Teff^{\max}$; $x_{\rm peak}$, $C_{\max}$: maximum on the stable branch; $x_z$: zero of $C$.}
\label{tab:published}
\end{table}

\subsection{Comparison}

\begin{figure}[ht]
\centering
\includegraphics[width=0.62\textwidth]{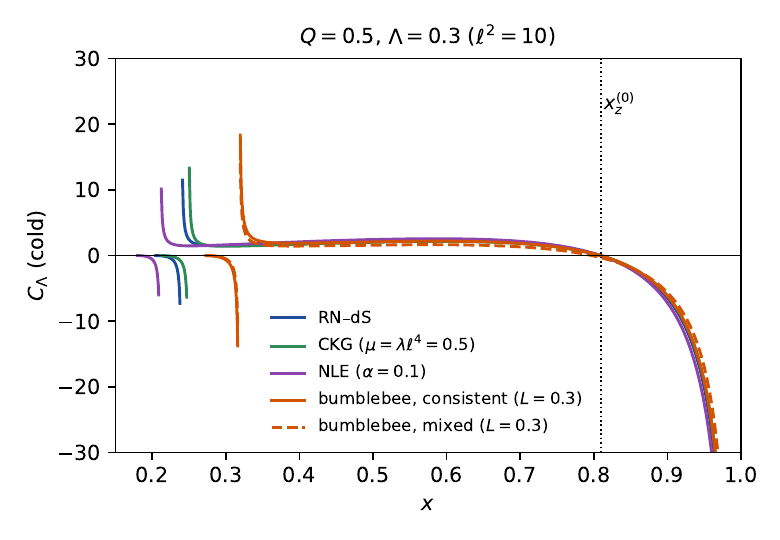}
\caption{Cold heat capacity $C_\Lambda$ at $\Lambda=0.3$ ($\ell^2=10$) for RN--dS ($Q=0.5$), Conformal Killing gravity ($q=0.5$, $\mu=\lambda\ell^4=0.5$), nonlinear electrodynamics ($q=0.5$, $\alpha=0.1$, $\Lambda_{\rm eff}=0.3$) and bumblebee gravity ($Q=0.5$, $L=0.3$) in the consistent normalization (ii) (solid) and the mixed normalization (iii) with $F=F_L$ (dashed). Each curve starts at its cold boundary. The dotted line marks the universal value $x_z^{(0)}$. Inset: the region $0.78<x<0.82$, where the zeros lie.}%Cold heat capacity $C_\Lambda$ at $Q=0.5$, $\Lambda=0.3$ for RN--dS, Conformal Killing gravity ($\lambda=0.005$) and nonlinear electrodynamics ($\alpha=0.1$), and bumblebee gravity ($L=0.3$, surface-gravity temperature, $F=F_L$), each plotted from its cold boundary. The dotted line marks the universal leading-order zero $x_z^{(0)}$ of the class.}
\label{fig:comparison}
\end{figure}

Fig.~\ref{fig:comparison} shows the cold heat capacity $C_\Lambda$ for the four spacetimes at the same charge $0.5$ ($Q$ for RN--dS and bumblebee gravity, $q$ for Conformal Killing gravity and nonlinear electrodynamics) and $\Lambda=0.3$ ($\ell^2=10$). The parameters are illustrative and differ from those of Tables~\ref{tab:peaks}--\ref{tab:published}: Conformal Killing gravity with $\mu=\lambda\ell^4=0.5$; nonlinear electrodynamics with $\alpha=0.1$ and $\Lambda_{\rm eff}=0.3$, i.e.\ $\ell^2=3/(\Lambda_{\rm eff}-\alpha^2)=10.34$; bumblebee gravity with $L=0.3$, i.e.\ $\Lambda_{\rm eff}=0.39$. RN--dS, Conformal Killing gravity and nonlinear electrodynamics differ in their cold boundary and in the position of the pole, which depend on $H$ and $k$ through $T_+$ and $T_c$. Their zeros, however, all lie between $0.808$ and $0.810$, close to the universal value $x_z^{(0)}=0.80953$ of Eq.~\eqref{eq:xzvalues}. The bumblebee curve in the consistent normalization has its zero at the same place, $0.809$; only the mixed normalization (iii) shifts it to $0.797$, as predicted by Eq.~\eqref{eq:FL}. The inset of Fig.~\ref{fig:comparison} enlarges the region of the zeros.

\section{Calibration dependence of published conclusions}
\label{sec:calibration}

\subsection{RN--dS}

\begin{figure}[ht]
\centering
\includegraphics[width=0.48\textwidth]{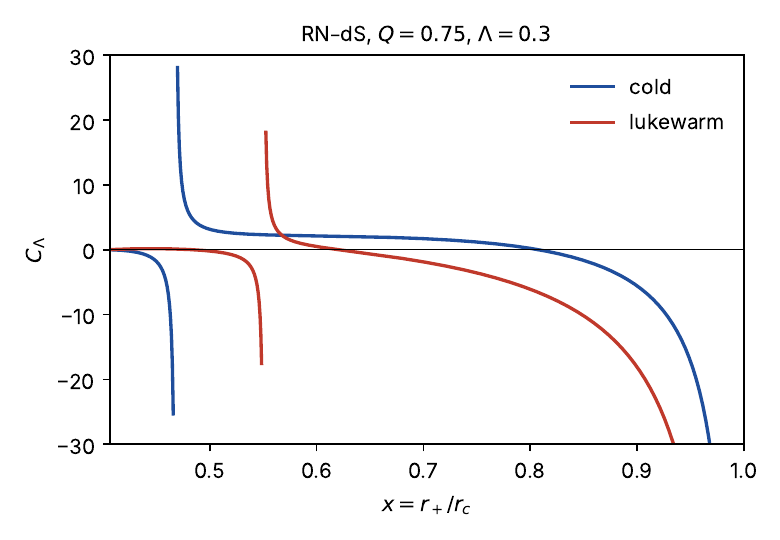}\hfill
\includegraphics[width=0.48\textwidth]{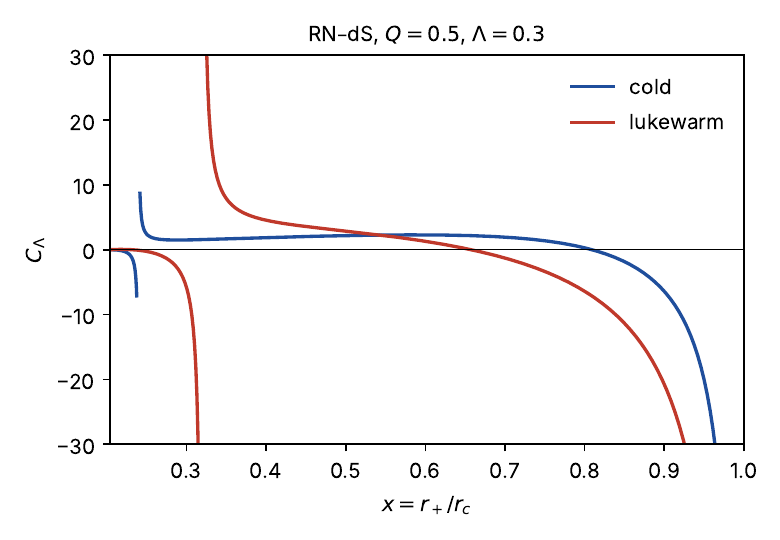}
\caption{Heat capacity $C_\Lambda$ of RN--dS at $\Lambda=0.3$ for $Q=0.75$ (left) and $Q=0.5$ (right), with the cold (blue) and lukewarm (red) calibrations, plotted from the cold boundary $x_A$. Insets: the narrow window $x_A<x<x_1$ where the lukewarm $C_\Lambda$ is positive and has its maximum.}
\label{fig:CLambda}
\end{figure}

We take RN--dS ($k=1$, $H=Q^2/r^2$) with $\Lambda=0.3$ and compute $C_\Lambda$ with both calibrations. This is the spacetime of Ref.~\cite{Zhen2025}, which introduced the cold calibration; its own parameters are treated in Section~\ref{sec:paper3luke}. For each charge, $r_c^2(x)$ is the cosmological root of Eq.~\eqref{eq:rcLambda}, $\Teff$ is given by Eq.~\eqref{eq:Teffcold} or~\eqref{eq:Teffluke}, and $S=\pi r_c^2F$ with the corresponding $F$. The physical domain starts at the cold boundary $x_A$, where $T_+=0$. The results are in Table~\ref{tab:RNdS} and Fig.~\ref{fig:CLambda}.

\begin{table}[ht]
\centering
\small
\renewcommand{\arraystretch}{1.2}
\begin{tabular}{lcc}
\toprule
 & $Q=0.75$ & $Q=0.5$ \\
\midrule
Cold boundary $x_A$ ($T_+=0$) & $0.40687$ & $0.20435$ \\
Pole of $C_\Lambda$, cold & $0.46787$ & $0.23969$ \\
Pole of $C_\Lambda$, lukewarm & $0.55056$ & $0.31970$ \\
Zero of $C_\Lambda$, cold & $0.80746$ & $0.80902$ \\
Upper zero of $C_\Lambda$, lukewarm & $0.62056$ & $0.65729$ \\
Lower zero $x_1$ of $C_\Lambda$, lukewarm & $0.48352$ & $0.23178$ \\
\midrule
Window with $C_\Lambda>0$, cold & $0.468$--$0.807$ & $0.240$--$0.809$ \\
Windows with $C_\Lambda>0$, lukewarm & $0.407$--$0.484$, $0.551$--$0.621$ & $0.204$--$0.232$, $0.320$--$0.657$ \\
Ratio of total widths (cold/lukewarm) & $2.32$ & $1.56$ \\
Maximum, cold: $x$, $C_\Lambda$ & none & $0.5873$, $2.283$ \\
Maximum, lukewarm: $x$, $C_\Lambda$ & $0.4455$, $0.154$ & $0.2184$, $0.0486$ \\
Crossing of the two curves & $0.5675$ & $0.5436$ \\
\bottomrule
\end{tabular}
\caption{Heat capacity $C_\Lambda$ of RN--dS at $\Lambda=0.3$ with the cold and lukewarm calibrations. With the lukewarm calibration $C_\Lambda>0$ on two windows, $x_A<x<x_1$ and from the pole to the upper zero; with the cold calibration only on the second.
}
\label{tab:RNdS}
\end{table}

\begin{figure}[ht]
\centering
\includegraphics[width=0.55\textwidth]{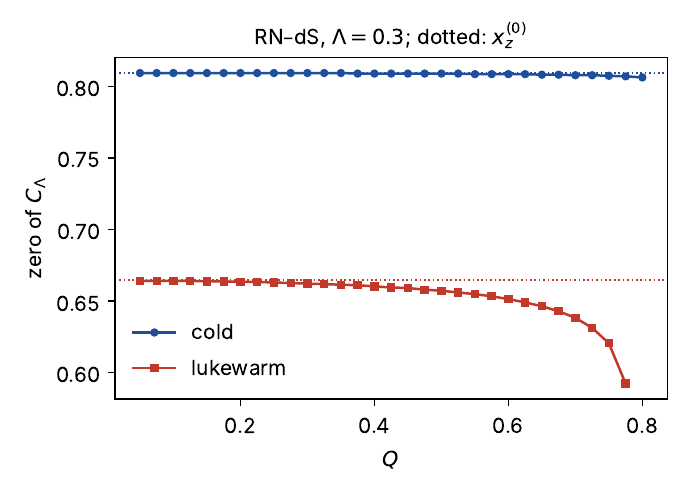}
\caption{Zero of $C_\Lambda$ at the upper end of the stable window, as a function of $Q$, for RN--dS at $\Lambda=0.3$ with the cold and lukewarm calibrations. Dotted lines: the universal values $x_z^{(0)}$ of Eq.~\eqref{eq:xzvalues}.}
\label{fig:xzQ}
\end{figure}

The following conclusions depend on the calibration.
\begin{enumerate}
\item \emph{Positive windows.} With the cold calibration $C_\Lambda>0$ on one window, from the pole to the zero near $0.81$. With the lukewarm calibration, it is positive on two:, a narrow one next to the cold boundary and one from the pole to the zero near $0.62$--$0.66$. The total width is larger with the cold calibration, by a factor that depends on $Q$: $2.32$ at $Q=0.75$ and $1.56$ at $Q=0.5$. 

\item \emph{Maximum.} At $Q=0.5$ the cold maximum lies at $x=0.587$ with $C_\Lambda=2.28$, while the only lukewarm maximum lies in the narrow window, at $x=0.218$ with $C_\Lambda=0.049$. At $Q=0.75$ the cold $C_\Lambda$ has no interior maximum, and the lukewarm one has its maximum at $x=0.446$, again in the narrow window.
\item \emph{Zero of the heat capacity.} The zero moves from near $0.81$ (cold) to near $0.62$--$0.66$ (lukewarm). As $Q$ decreases, both approach the universal values~\eqref{eq:xzvalues}, as shown in Fig.~\ref{fig:xzQ}. The cold zero barely moves with $Q$, while the lukewarm zero is more sensitive to the charge.
\item \emph{Magnitude.} Between the lukewarm pole and the crossing point, the lukewarm $C_\Lambda$ is larger, so the calibrations differ in the size of $C$ as well as in where it changes sign.
\item \emph{Small black holes.} Only the lukewarm $F$ contains the black-hole area at leading order, Eqs.~\eqref{eq:Fcold0}--\eqref{eq:Fluke0}.
\item \emph{Nariai limit.} The coefficient of the pole of $F$ doubles, from $2/15$ to $4/15$, Eq.~\eqref{eq:Nariai}.
\item \emph{Lukewarm states.} With the cold calibration $\Teff=T(1+x^{-4})$ at a lukewarm state, Eq.~\eqref{eq:conflict}, while the lukewarm calibration sets $\Teff=T$ by construction.
\end{enumerate}
\subsection{Published Schottky-like peaks under the other calibration}
\label{sec:paper3luke}

We now recompute published heat capacities with the other calibration, i.e.\ with $\Teff=(x^4T_c+T_+)/(1+x^4)$ and $F=1+x^2+f_0$ in place of $\Teff=T_c+T_+/x^4$ and $F_{\rm cold}$, and nothing else changed. We start with the two studies that established the Schottky interpretation with the cold calibration. Ref.~\cite{Zhen2025} studied RN--dS with $Q=1$ and $\ell^2=12\alpha Q^2$, $\alpha=30$, $50$, $100$. Ref.~\cite{KRdS2024} studied the Kalb--Ramond--dS spacetime, $f=k-2M/r+Q^2/[(1-a)^2r^2]-r^2/[\ell^2(1-a)]$ with $k=1/(1-a)$, for $Q^2/(1-a)=1$ and $\ell^2=12\gamma Q^2/(1-a)$, $\gamma=30$, $70$, $100$. This metric is $k$ times an RN--dS metric with unit charge, and an overall factor of $f$ rescales $\Teff$ but cancels in $C_\Lambda$, so the results for $\gamma$ equal those of Ref.~\cite{Zhen2025} for $\alpha=\gamma$. With the cold calibration, we reproduce the published curves, for example the maxima $C_{\max}\simeq89$, $149$, $298$ at $x\simeq0.58$ in Fig.~8 of Ref.~\cite{Zhen2025}, and the coefficient $C_{\max}\simeq2.9\gamma$ of Ref.~\cite{KRdS2024}, for which we find $2.96$--$2.98$. Table~\ref{tab:external} lists the results together with those for the bumblebee black hole with the parameters of Ref.~\cite{Paper1}.

\begin{table}[ht]
\centering
\small
\setlength{\tabcolsep}{4pt}
\renewcommand{\arraystretch}{1.2}
\begin{tabular}{ll|cc|cc|cc|cc}
\toprule
 & & & & \multicolumn{2}{c|}{cold} & \multicolumn{2}{c|}{lukewarm, $x_f<x<x_z$} & \multicolumn{2}{c}{lukewarm, $x_A<x<x_1$}\\
Spacetime & parameter & $x_A$ & $x_1$ & $x_{\rm peak}$ & $C_{\max}$ & $x_{\rm peak}$ & $C_{\max}$ & $x_{\rm peak}$ & $C_{\max}$\\
\midrule
RN--dS~\cite{Zhen2025} & $\alpha=30$  & 0.05605 & 0.05892 & 0.58239 & 88.886 & 0.28129 & 157.66 & 0.05750 & 0.0782\\
                       & $\alpha=50$  & 0.04275 & 0.04445 & 0.58210 & 148.68 & 0.29030 & 261.43 & 0.04361 & 0.0614\\
                       & $\alpha=100$ & 0.02979 & 0.03064 & 0.58188 & 298.16 & 0.29490 & 521.62 & 0.03022 & 0.0440\\
KR--dS~\cite{KRdS2024} & $\gamma=70$  & 0.03585 & 0.03706 & 0.58198 & 208.47 & 0.29309 & 365.46 & 0.03646 & 0.0523\\
\midrule
bumblebee~\cite{Paper1} & $\zeta=30$  & 0.08815 & 0.09481 & 0.58333 & 62.742 & none    & --     & 0.09154 & 0.1863\\
                        & $\zeta=70$  & 0.05545 & 0.05825 & 0.58238 & 148.17 & 0.28186 & 262.68 & 0.05687 & 0.1265\\
                        & $\zeta=140$ & 0.03844 & 0.03983 & 0.58202 & 297.66 & 0.29214 & 522.34 & 0.03914 & 0.0910\\
\bottomrule
\end{tabular}
\caption{Canonical heat capacity of published Schottky-type studies with the cold calibration (as published) and the lukewarm calibration. RN--dS: $Q=1$, $\ell^2=12\alpha Q^2$. KR--dS: $Q^2/(1-a)=1$, $\ell^2=12\gamma Q^2/(1-a)$; the rows for $\gamma=30$ and $100$ equal those for $\alpha=30$ and $100$. Bumblebee: $Q=1$, $L=0.4$, consistent normalization (ii). $x_A$: cold boundary; $x_1$: lower zero of the lukewarm $C_\Lambda$.}
\label{tab:external}
\end{table}

For Refs.~\cite{Zhen2025,KRdS2024} the lukewarm calibration moves the maximum from $x\simeq0.58$ to $x\simeq0.28$--$0.29$ and raises it by a factor of $1.75$--$1.77$; a second, much lower maximum appears in the narrow window next to the cold boundary. Any quantity matched to the peak, such as the number of microstates estimated from $C_{\max}$, changes accordingly.

The same test applies to our own earlier work. The conclusions of Ref.~\cite{Paper3} (see also Ref.~\cite{Zhen2024CQG}, where the Schottky anomaly of a nonlinearly charged dS spacetime was studied with effective quantities of the same type) were obtained with the cold calibration, and Section~\ref{sec:examples} shows that the general formulas reproduce them exactly. Table~\ref{tab:paper3luke} gives them for the same five configurations with both calibrations.

\begin{table}[ht]
\centering
\small
\setlength{\tabcolsep}{4pt}
\renewcommand{\arraystretch}{1.2}
\begin{tabular}{cc|c|cccc|c|c}
\toprule
$\alpha$ & $\omega$ & calibration & $x_1$ & $x_f$ & $x_{\rm peak}$ & $C_{\max}$ & $(\hat C/Nk)_{x_A}$ & $N_b\ln2/S_{\rm eff}$ \\
\midrule
0.25 & 1.2 & cold & -- & 0.36166 & 0.58205 & 2.04136 & 0.196612 & 0.633 \\
     &     & luke & 0.36028 & 0.44977 & $0.33635^{*}$ & $0.09248^{*}$ & $2.0\times10^{-43}$ & -- \\
0.35 & 1.2 & cold & -- & 0.33887 & 0.58529 & 2.00804 & 0.196612 & 0.630 \\
     &     & luke & 0.33590 & 0.42677 & $0.31405^{*}$ & $0.08143^{*}$ & $1.9\times10^{-57}$ & -- \\
0.45 & 1.2 & cold & -- & 0.32873 & 0.58615 & 1.90691 & 0.196612 & 0.629 \\
     &     & luke & 0.32513 & 0.41636 & $0.30418^{*}$ & $0.07337^{*}$ & $2.2\times10^{-65}$ & -- \\
0.35 & 1.6 & cold & -- & 0.28784 & 0.58765 & 2.53951 & 0.196612 & 0.625 \\
     &     & luke & 0.28203 & 0.37321 & $0.26462^{*}$ & $0.07704^{*}$ & $1.7\times10^{-114}$ & -- \\
0.35 & 2.0 & cold & -- & 0.25812 & 0.58761 & 2.99871 & 0.196612 & 0.622 \\
     &     & luke & 0.25096 & 0.34055 & $0.23606^{*}$ & $0.07395^{*}$ & $3.4\times10^{-180}$ & -- \\
\bottomrule
\end{tabular}
\caption{Canonical results of Ref.~\cite{Paper3} ($q=0.75$, $\epsilon=1$) with the cold calibration (as published) and recomputed with the lukewarm calibration. The cold boundary $x_A$ is the same for both (Table~\ref{tab:published}). $x_1$: lower zero of the lukewarm $C_\Lambda$; $x_f$: pole. $^{*}$The lukewarm $C_\Lambda$ has no maximum on its main window $x_f<x<x_z$ ($x_z=0.646$--$0.656$); the maximum listed lies in the narrow window $x_A<x<x_1$. $(\hat C/Nk)_{x_A}$: two-level heat capacity at the cold boundary; $N_b=C_{\max}/0.196612$ and $S_{\rm eff}=\pi r_c^2F$ at the peak, as in Ref.~\cite{Paper3}.}
\label{tab:paper3luke}
\end{table}

Three conclusions of Ref.~\cite{Paper3} change.
\begin{enumerate}
\item \emph{The maximum moves and drops.} With the lukewarm calibration $C_\Lambda$ has no maximum on its main window, where it decreases monotonically from the pole to the zero. Its only maximum lies in the narrow window next to the cold boundary, at $x=0.236$--$0.336$ instead of $x\simeq0.58$, and is lower than the cold maximum by a factor of $22$ to $41$ as discussed in Section~\ref{sec:general}; this window is a feature of the lukewarm entropy function near the cold boundary.
\item \emph{The cold-boundary value is not parameter-independent.} In Ref.~\cite{Paper3} the two-level heat capacity at the cold boundary takes the value $(\epsilon/2)^2\,\mathrm{sech}^2(\epsilon/2)=0.196612$ for all parameters. At $x_A$ one has $T_+=0$, so $\Delta/T=-\epsilon\,T_c/\Teff$. With the cold calibration $\Teff=T_c$ there by construction, which gives $\Delta/T=-\epsilon$ exactly. With the lukewarm calibration, Eq.~\eqref{eq:Teffluke} gives $\Teff=x_A^4T_c/(1+x_A^4)$ and hence $\Delta/T=-\epsilon(1+x_A^4)/x_A^4$, which depends on the parameters through $x_A$ and makes $(\hat C/Nk)_{x_A}$ smaller than $10^{-42}$ in all five cases. The parameter independence of $0.196612$ is therefore the cold calibration condition~\eqref{eq:coldcond} itself, not an independent property of the spacetime.
\item \emph{The entropy-bound ratio.} The ratio $N_b\ln2/S_{\rm eff}\simeq0.62$--$0.63$ relies on both the cold-boundary normalization and the cold maximum at $x\simeq0.58$, neither of which survives the change of calibration. The bound $N_b\ln2<S_{\rm eff}$ found in Ref.~\cite{Paper3} is therefore a statement about the cold calibration.
\end{enumerate}
For Conformal Killing gravity with the parameters of Ref.~\cite{Paper2}, the lukewarm maximum on the main window lies at $x_{\rm peak}=0.32$--$0.36$ instead of $0.59$--$0.61$, with a different height and a temperature lower by a factor of $3.4$--$3.7$ (Table~\ref{tab:peaks}). For the bumblebee black hole of Ref.~\cite{Paper1}, it lies at $x=0.282$ and $0.292$ for $\zeta=70$ and $140$, while for $\zeta=30$ the only lukewarm maximum is in the narrow window, at $x=0.0915$ (Table~\ref{tab:external}). In all cases examined, the position and the height of the Schottky-like maximum, and any counting of microstates matched to it, depend on the calibration.

None of this shows that these results are incorrect within the cold calibration, which Section~\ref{sec:examples} confirms to all quoted digits. It shows that the Schottky-like interpretation, and the counting built on it, depend on the calibration choice. 

\section{Conclusions}
\label{sec:conclusions}

We have shown what the effective two-horizon method depends on, for the class~\eqref{eq:metric} of spacetimes $f(r)=k-2M/r-\Lambda r^2/3+H(r)$. 
The main physical consequence is that the Schottky-like peaks of the canonical heat capacity found with this method, and the counting of horizon microstates matched to them, are not properties of the spacetime alone: their position and their height depend on the calibration that fixes $F$. Section~\ref{sec:paper3luke} shows this for published studies~\cite{Zhen2025,KRdS2024,Paper1,Paper2,Paper3}.

\emph{What the method depends on.} The entropy function $F(x)$ is fixed by the calibration alone. It is the same for every spacetime in the class, for both calibrations in use (Theorem~\ref{thm:universal}, Corollary~\ref{cor:luke}), and for every normalization of temperature, energy and entropy in which each horizon obeys its own first law, $dE=T_+dS_+$ at $r_+$ and $dE=-T_c\,dS_c$ at $r_c$ (Section~\ref{sec:normalization}). With a mixed normalization, such as the Bousso--Hawking temperature with the mass as energy, $F$ depends on the metric. The effective temperature always has the form $\Teff=\Omega(x)(T_c+T_+/x^4)$, so the metric enters only through the two horizon temperatures, and the calibration only through $\Omega$ (Theorem~\ref{thm:structure}). For the cold calibration, $\Teff=T_c+T_+/x^4$ exactly. The class ends where (A1), (A2) or (A4) fails, and we gave a counterexample for each (Sections~\ref{sec:boundary} and~\ref{sec:normalization}). The universality is a consistency property of the entropy ansatz; what it adds is the class-wide statement, its boundary, and the normalization condition $as=b$.

\emph{What changes with the calibration.} The cold and lukewarm calibrations are incompatible for every spacetime in the class (Corollary~\ref{cor:conflict}). They give different entropy functions, Eq.~\eqref{eq:DeltaF}, and hence different heat capacities, positive windows, small-$x$ behaviour and Nariai limits (Section~\ref{sec:calibration}). With the lukewarm calibration, $C_\Lambda$ is positive on two windows, a narrow one next to the cold boundary and a main one below $x\simeq0.66$, while with the cold calibration it is positive on one window ending near $0.81$. Results obtained with one calibration therefore cannot be compared directly with results obtained with the other. For RN--dS~\cite{Zhen2025}, Kalb--Ramond--dS~\cite{KRdS2024}, Conformal Killing gravity~\cite{Paper2} and bumblebee gravity~\cite{Paper1}, the lukewarm maximum lies at $x\simeq0.28$--$0.36$ instead of $x\simeq0.58$--$0.61$. For the nonlinearly charged dS black hole~\cite{Paper3}, the lukewarm $C_\Lambda$ has no maximum on its main window; its only maximum lies in the narrow window next to the cold boundary and is lower by a factor of $22$--$41$. The parameter-independent cold-boundary value $0.196612$ used to normalize the two-level counting in Ref.~\cite{Paper3} turns out to be the cold calibration condition itself (Section~\ref{sec:paper3luke}).

\emph{Results for the parameters of Refs.~\cite{Paper1,Paper2}.} For the parameters of Ref.~\cite{Paper2} the general formulas place the heat-capacity maximum of Conformal Killing gravity at $x=0.59$--$0.61$ (Section~\ref{sec:CKG}). For bumblebee gravity the physical temperature is the surface-gravity one; with the consistent energy and entropy the black hole belongs to the class, the positions in $x$ found in Ref.~\cite{Paper1} are unchanged, and its temperatures and heat capacities are rescaled by $1/\sqrt{1+L}$ and $1+L$ (Section~\ref{sec:bumblebee}).

\emph{What the method does not settle.} The theorems show what the method depends on, not that it is correct. They do not decide which calibration, if either, is physically preferred. Neither calibration recovers the de~Sitter temperature in the small-black-hole limit, and the effective entropy is subadditive only below $x=0.9411$ (cold) or $x=0.7696$ (lukewarm) (Section~\ref{sec:limits}). Neither calibration reproduces the entropy $\pi r_c^2(1+x^2)$ obtained from the on-shell Euclidean action of RN--dS~\cite{MvdSV2024}, and relating the effective method to the ensembles with a boundary of Refs.~\cite{GT2003,BJ2022} is an open problem. %REV3 M4
The theorems also do not explain the pole of $F$ at the Nariai limit, which produces a divergence $C_\Lambda\propto-1/\Teff$ of calibration-independent strength (Proposition~\ref{prop:Nariai}). The fixed weighting $1:x^{-4}$ of the two horizon temperatures in Eq.~\eqref{eq:main} raises the question of whether a two-level description of the horizon pair can be derived without relying on the shape of the heat-capacity curve. We leave this as an open question.

\emph{Outlook.} Sections~\ref{sec:boundary} and~\ref{sec:normalization} identify where the universality ends, and each boundary is a natural extension. Non-area entropies such as the Barrow, R\'enyi and Tsallis forms~\cite{Barrow2020,CI2016,TC2013}, and the Gauss--Bonnet entropy~\cite{Guo2020GB}, change the scaling of $S$ with $r_c$, so that no function of $x$ alone can satisfy the calibration; how the method must be modified in that case is open. In $D$ dimensions the same steps go through with $S\propto r_c^{D-2}$ and $V\propto r^{D-1}$, and give a different universal $F$ for each $D$. Finally, a physical criterion for choosing between the calibrations, independent of the shape of the resulting heat capacity, would remove the main ambiguity identified here.

\section*{Acknowledgments}
The present research work was funded by Princess Nourah bint Abdulrahman University Researchers Supporting
Project number (PNURSP2026R59), Princess Nourah bint Abdulrahman University, Riyadh, Saudi Arabia.

\section*{Use of AI}
The authors used different chatbots to enhance the linguistic clarity and grammar of the manuscript. AI was also used to improve the visual resolution and formatting of figures. All substantive intellectual content, analysis, and conclusions remain the work of the authors.

\section*{Data Availability Statement}
No data were generated or analyzed in this study.

%\appendix
\section{Symbolic and numerical verification}
\label{app:verification}

We checked every identity symbolically and, where indicated, numerically to 30 digits in Maple and independently in Python (SymPy, mpmath). Unless stated otherwise, the numerical checks use $(r_+,r_c)=(1,3)$.

\begin{table}[ht]
\centering
\footnotesize
\renewcommand{\arraystretch}{1.2}
\begin{tabular}{lll}
\toprule
\VA{Check} & \VB{Result} & \VC{Script} \\
\midrule
\VA{Eq.~\eqref{eq:coldODE} solved by Eq.~\eqref{eq:Fcold}; lukewarm ODE solved by Eq.~\eqref{eq:f0}; Eq.~\eqref{eq:DeltaF}} & \VB{$0$ (symbolic)} & \VC{Maple, SymPy} \\
\VA{Eqs.~\eqref{eq:Mrcx}--\eqref{eq:Lrcx} and Lemma~\ref{lem:dM} for arbitrary $H$, $k$, $F$} & \VB{$0$ (symbolic)} & \VC{Maple, SymPy} \\
\VA{Eq.~\eqref{eq:Teffnorm} for arbitrary $a$, $b$, $s$; Eq.~\eqref{eq:singlefirstlaw}} & \VB{$0$; factor $(as-b)$} & \VC{\texttt{universality\_checks.py}} \\
\VA{Single-horizon first laws, RN--dS, $(r_+,r_c,Q)=(1,3,0.4)$} & \VB{$dM=T_+dS_+$; $dM=-T_c\,dS_c$} & \VC{\texttt{universality\_checks.py}} \\
\VA{Eq.~\eqref{eq:Teffcold}: RN--dS ($Q=0.4$), CKG ($Q=0.4$, $\lambda=0.01$), NLE ($q=0.4$, $\alpha=0.1$)} & \VB{$0$ (symbolic); $<10^{-23}$ (numerical)} & \VC{Maple, \texttt{universality\_checks.py}} \\
\VA{Eq.~\eqref{eq:Teffluke}, RN--dS, $Q=0.4$} & \VB{$0$; $10^{-31}$} & \VC{Maple} \\
\VA{$\Peff$, Eq.~\eqref{eq:Peff}, against $-(\partial M/\partial V)_S$; Smarr~\eqref{eq:smarr}, RN--dS and CKG} & \VB{$<10^{-25}$; $<10^{-23}$} & \VC{\texttt{universality\_checks.py}} \\
\VA{Counter-examples: (A1) with $\beta=0.1$; mass term with $\gamma$} & \VB{ratio $1.061$; ratio $1/(1+\gamma)$} & \VC{Maple, \texttt{universality\_checks.py}} \\
\VA{Eq.~\eqref{eq:FL}: ODE, $F_L(0)=1$, $F_L|_{L=0}=F_{\rm cold}$; zeros for $L=0.1,0.3,0.5,1$} & \VB{$0$, $1$, $0$; Section~\ref{sec:normalization}} & \VC{SymPy} \\
\VA{Bousso--Hawking factor $f(r_0)^{-1/2}$ on the cold lines, $x=0.3,0.5,0.7$} & \VB{differs for RN--dS and CKG} & \VC{\texttt{universality\_checks.py}} \\
\VA{Cold and lukewarm lines of RN--dS, Eq.~\eqref{eq:coldline}; positivity on $(0,1)$} & \VB{exact; yes} & \VC{\texttt{universality\_checks.py}} \\
\VA{Eq.~\eqref{eq:xzvalues}} & \VB{$0.8095321656$, $0.6644953859$} & \VC{SymPy} \\
\VA{Eqs.~\eqref{eq:Fcold0}--\eqref{eq:Fluke0}, \eqref{eq:Nariai}; subadditivity crossings} & \VB{series; $2/15$, $4/15$; $0.9411$, $0.7696$} & \VC{\texttt{universality\_checks.py}} \\
\VA{Eq.~\eqref{eq:xzexact} at $x_z$ and Table~\ref{tab:chi} (11 configurations)} & \VB{residual $<10^{-25}$} & \VC{\texttt{universality\_checks.py}} \\
\VA{Zero $x_z$ of the cold $C_\Lambda$, RN--dS, $\Lambda=0.3$, $Q=0.5$--$0.86$} & \VB{$-0.06\%$ to $-0.76\%$} & \VC{\texttt{positive\_windows.py}} \\
\VA{Prop.~\ref{prop:Nariai}, RN--dS, $1-x=10^{-5}$} & \VB{$-1.28220$, $-2.56445$; $C\Teff=-0.0975880$} & \VC{Maple} \\
\VA{Tables~\ref{tab:peaks}, \ref{tab:paper1}: maxima located by bisection of $dC/dx$} & \VB{as tabulated} & \VC{\texttt{heat\_capacity\_examples.py}, \texttt{universality\_checks.py}} \\
\VA{CKG, Sec.~\ref{sec:CKG}: fixed $\Lambda$ ($\eta=30$--$140$); other roots of the cubic; fixed $V$} & \VB{$x_{\rm peak}=0.595$--$0.609$; no maximum; no maximum} & \VC{\texttt{ckg\_heat\_capacity.py}} \\
\VA{Table~\ref{tab:published} against Table~IV of Ref.~\cite{Paper3} (35 values)} & \VB{max.\ rel.\ diff.\ $4\times10^{-5}$} & \VC{\texttt{nle\_heat\_capacity.py}} \\
\VA{Positive windows and maxima of $C_\Lambda$, both calibrations: Tables~\ref{tab:peaks}, \ref{tab:RNdS}, \ref{tab:external}, \ref{tab:paper3luke}} & \VB{as tabulated} & \VC{\texttt{positive\_windows.py}} \\
\VA{Table~\ref{tab:paper3luke}: $(\hat C/Nk)_{x_A}$ and $N_b\ln2/S_{\rm eff}$} & \VB{as tabulated} & \VC{\texttt{nle\_lukewarm.py}} \\
\VA{Table~\ref{tab:RNdS}, Figs.~\ref{fig:FOmega}--\ref{fig:xzQ}} & \VB{as shown} & \VC{\texttt{rnds\_heat\_capacity.py}, \texttt{all\_figures.py}} \\
\bottomrule
\end{tabular}
\caption{Checks of the results. ``Numerical'' residuals are absolute differences at 30 significant digits.}
\label{tab:verification}
\end{table}


\begin{thebibliography}{99}

\bibitem{GH1977}
G.~W.~Gibbons, S.~W.~Hawking,
``Cosmological event horizons, thermodynamics, and particle creation,''
Phys.\ Rev.\ D \textbf{15}, 2738 (1977).

\bibitem{Cai2002}
R.-G.~Cai,
``Cardy--Verlinde formula and thermodynamics of black holes in de Sitter spaces,''
Nucl.\ Phys.\ B \textbf{628}, 375 (2002), arXiv:hep-th/0112253.

\bibitem{Sekiwa2006}
Y.~Sekiwa,
``Thermodynamics of de Sitter black holes: thermal cosmological constant,''
Phys.\ Rev.\ D \textbf{73}, 084009 (2006), arXiv:hep-th/0602269.

\bibitem{Urano2009}
M.~Urano, A.~Tomimatsu, H.~Saida,
``Mechanical first law of black hole spacetimes with cosmological constant and its application to Schwarzschild--de Sitter spacetime,''
Class.\ Quantum Grav.\ \textbf{26}, 105010 (2009), arXiv:0903.4230.

\bibitem{Dolan2013}
B.~P.~Dolan, D.~Kastor, D.~Kubiz\v{n}\'ak, R.~B.~Mann, J.~Traschen,
``Thermodynamic volumes and isoperimetric inequalities for de Sitter black holes,''
Phys.\ Rev.\ D \textbf{87}, 104017 (2013), arXiv:1301.5926.

\bibitem{KS2016}
D.~Kubiz\v{n}\'ak, F.~Simovic,
``Thermodynamics of horizons: de Sitter black holes and reentrant phase transitions,''
Class.\ Quantum Grav.\ \textbf{33}, 245001 (2016), arXiv:1507.08630.

\bibitem{KMT2017}
D.~Kubiz\v{n}\'ak, R.~B.~Mann, M.~Teo,
``Black hole chemistry: thermodynamics with Lambda,''
Class.\ Quantum Grav.\ \textbf{34}, 063001 (2017), arXiv:1608.06147.

\bibitem{GT2003}
A.~Gomberoff, C.~Teitelboim,
``de Sitter black holes with either of the two horizons as a boundary,''
Phys.\ Rev.\ D \textbf{67}, 104024 (2003), arXiv:hep-th/0302204.

\bibitem{BJ2022}
B.~Banihashemi, T.~Jacobson,
``Thermodynamic ensembles with cosmological horizons,''
J.\ High Energy Phys.\ \textbf{07}, 042 (2022), arXiv:2204.05324.

\bibitem{BJSV2023}
B.~Banihashemi, T.~Jacobson, A.~Svesko, M.~R.~Visser,
``The minus sign in the first law of de Sitter horizons,''
J.\ High Energy Phys.\ \textbf{01}, 054 (2023), arXiv:2208.11706.

\bibitem{MvdSV2023}
E.~K.~Morvan, J.~P.~van~der~Schaar, M.~R.~Visser,
``On the Euclidean action of de Sitter black holes and constrained instantons,''
SciPost Phys.\ \textbf{14}, 022 (2023), arXiv:2203.06155.

\bibitem{MvdSV2024}
E.~Morvan, J.~P.~van~der~Schaar, M.~R.~Visser,
``Action, entropy and pair creation rate of charged black holes in de Sitter space,''
J.\ High Energy Phys.\ \textbf{08}, 207 (2024), doi:10.1007/JHEP08(2024)207.

\bibitem{Ma2014}
M.-S.~Ma, H.-H.~Zhao, L.-C.~Zhang, R.~Zhao,
``Existence condition and phase transition of Reissner--Nordstr\"om--de Sitter black hole,''
Int.\ J.\ Mod.\ Phys.\ A \textbf{29}, 1450050 (2014), arXiv:1312.0731.

\bibitem{LiMaMa2017}
H.-F.~Li, M.-S.~Ma, Y.-Q.~Ma,
``Thermodynamic properties of black holes in de Sitter space,''
Mod.\ Phys.\ Lett.\ A \textbf{32}, 1750017 (2017), arXiv:1605.08225.

\bibitem{ZZM2016}
L.-C.~Zhang, R.~Zhao, M.-S.~Ma,
``Entropy of Reissner--Nordstr\"om--de Sitter black hole,''
Phys.\ Lett.\ B \textbf{761}, 74 (2016), arXiv:1610.09886.

\bibitem{MZC2018}
M.-S.~Ma, R.~Zhao, Y.-S.~Cao,
``Thermodynamics of de Sitter black holes in massive gravity,''
Commun.\ Theor.\ Phys.\ \textbf{69}, 544 (2018), doi:10.1088/0253-6102/69/5/544, arXiv:1708.01520.

\bibitem{DLZ2022}
Y.-Z.~Du, H.-F.~Li, R.~Zhao,
``Overview of thermodynamical properties for Reissner--Nordstr\"om--de Sitter spacetime in induced phase space,''
Eur.\ Phys.\ J.\ C \textbf{82}, 850 (2022).

\bibitem{Sereewat2026}
P.~Sereewat, D.~Senjaya, P.~Burikham,
``Thermodynamics and geometrical optics of Reissner--Nordstr\"om de Sitter black holes in noncommutative geometry,''
arXiv:2603.20824 (2026).

\bibitem{ZhaoZhang2018}
R.~Zhao, L.-C.~Zhang,
``Entropy of higher-dimensional charged de Sitter black holes and phase transition,''
Commun.\ Theor.\ Phys.\ \textbf{70}, 578 (2018), arXiv:1710.07225.

\bibitem{Zhen2025}
H.-L.~Zhen, Y.-B.~Ma, H.-F.~Li, L.-C.~Zhang, Y.-Z.~Du,
``Schottky anomaly of Reissner--Nordstr\"om--de Sitter spacetime,''
Chin.\ Phys.\ C \textbf{49}, 035105 (2025), arXiv:2406.05406.

\bibitem{KRdS2024}
Y.~Ma, S.~Zheng, H.~Li, B.~Li,
``Schottky anomaly of the Kalb--Ramond--de Sitter spacetime,''
Nucl.\ Phys.\ B \textbf{1009}, 116732 (2024).

\bibitem{MaZhao2025}
Y.~Ma, H.~Zhao,
``Schottky anomaly of a perturbed Reissner--N\"ordstrom de Sitter spacetime,''
Eur.\ Phys.\ J.\ C \textbf{85}, 1302 (2025), doi:10.1140/epjc/s10052-025-15022-y.

\bibitem{Paper1}
N.~Alessa, M.~U.~Shahzad, A.~Ashfaq,
``Discrete microstate spectra and Schottky anomaly in charged de Sitter black holes via bumblebee gravity,''
Nucl.\ Phys.\ B \textbf{1027}, 117477 (2026), doi:10.1016/j.nuclphysb.2026.117477.
\bibitem{Paper2}
N.~Alessa, M.~U.~Shahzad, A.~Ashfaq,
``Schottky anomaly and microstructure of de Sitter black holes in Conformal Killing Gravity,''
Int.\ J.\ Mod.\ Phys.\ A \textbf{41}, 2650177 (2026), doi:10.1142/S0217751X26501770.

\bibitem{Paper3}
N.~Alessa, A.~Ashfaq, M.~U.~Shahzad,
``Schottky peaks and the entropy bound: counting horizon-pair degrees of freedom in nonlinearly charged de Sitter black holes,''
submitted to Mod.\ Phys.\ Lett.\ A (2026). %CHECK title exactly as in the version under review at MPLA

\bibitem{Johnson2020}
C.~V.~Johnson,
``Specific heats and Schottky peaks for black holes in extended thermodynamics,''
Class.\ Quantum Grav.\ \textbf{37}, 054003 (2020), arXiv:1905.00539.

\bibitem{JohnsondS}
C.~V.~Johnson,
``de Sitter black holes, Schottky peaks, and continuous heat engines,''
arXiv:1907.05883 (2019).

\bibitem{Dinsmore2020}
J.~Dinsmore, P.~Draper, D.~Kastor, Y.~Qiu, J.~Traschen,
``Schottky anomaly of de Sitter black holes,''
Class.\ Quantum Grav.\ \textbf{37}, 054001 (2020), arXiv:1907.00248.

\bibitem{Ma2019CSQ}
Y.~Ma, Y.~Zhang, R.~Zhao, S.~Cao, T.~Liu, S.~Geng, Y.~Liu, Y.~Huang,
``Phase transitions and entropy force of charged de Sitter black holes with cloud of string and quintessence,''
arXiv:1907.11870 (2019).

\bibitem{Zhang2022Q}
Y.~Zhang, Y.-B.~Ma, Y.-Z.~Du, H.-F.~Li, L.-C.~Zhang,
``Thermodynamics of the Reissner--Nordstr\"om--de Sitter spacetime with quintessence,''
Eur.\ Phys.\ J.\ C \textbf{82}, 770 (2022), doi:10.1140/epjc/s10052-022-10729-8, arXiv:2204.04470.

\bibitem{Du2024KR}
Y.-Z.~Du, H.-F.~Li, Y.-B.~Ma, Q.~Gu,
``Phase structure and optical properties of the de Sitter spacetime with KR field based on the Lyapunov exponent,''
Eur.\ Phys.\ J.\ C \textbf{85}, 78 (2025), doi:10.1140/epjc/s10052-025-13809-7, arXiv:2403.20083.

\bibitem{Zhen2024CQG}
H.-L.~Zhen, Y.-Z.~Du, H.-F.~Li, L.-C.~Zhang, Y.-B.~Ma,
``Non-linear charged dS spacetime and its thermodynamics and Schottky anomaly,''
Class.\ Quantum Grav.\ \textbf{41}, 225018 (2024), doi:10.1088/1361-6382/ad828e.

\bibitem{Guo2020GB}
X.-Y.~Guo, Y.~Gao, H.-F.~Li, R.~Zhao,
``The entropic force between two horizons of charged Gauss--Bonnet black hole in de Sitter spacetime,''
arXiv:2007.03284 (2020).

\bibitem{Barrow2020}
J.~D.~Barrow,
``The area of a rough black hole,''
Phys.\ Lett.\ B \textbf{808}, 135643 (2020), arXiv:2004.09444.

\bibitem{TC2013}
C.~Tsallis, L.~J.~L.~Cirto,
``Black hole thermodynamical entropy,''
Eur.\ Phys.\ J.\ C \textbf{73}, 2487 (2013).

\bibitem{CI2016}
V.~G.~Czinner, H.~Iguchi,
``R\'enyi entropy and the thermodynamic stability of black holes,''
Phys.\ Lett.\ B \textbf{752}, 306 (2016), arXiv:1511.06963.

\bibitem{Hairy5D}
S.-X.~Bao, R.~Zhao, Y.-Z.~Du,
``Schottky anomaly of five-dimensional de Sitter hairy spacetime based on effective thermodynamic quantities,''
arXiv:2511.07873 (2025).

\bibitem{Visser1992}
M.~Visser,
``Dirty black holes: thermodynamics and horizon structure,''
Phys.\ Rev.\ D \textbf{46}, 2445 (1992), arXiv:hep-th/9203057.

\bibitem{KanziSakalli2019}
S.~Kanzi, \.I.~Sakall\i,
``GUP modified Hawking radiation in bumblebee gravity,''
Nucl.\ Phys.\ B \textbf{946}, 114703 (2019), doi:10.1016/j.nuclphysb.2019.114703, arXiv:1905.00477.

\bibitem{An2024}
Y.-S.~An,
``Notes on thermodynamics of Schwarzschild-like bumblebee black hole,''
Phys.\ Dark Univ.\ (2024) 101520, doi:10.1016/j.dark.2024.101520, arXiv:2401.15430. 

\bibitem{BH1996}
R.~Bousso, S.~W.~Hawking,
``Pair creation of black holes during inflation,''
Phys.\ Rev.\ D \textbf{54}, 6312 (1996), arXiv:gr-qc/9606052.

\bibitem{Romans1992}
L.~J.~Romans,
``Supersymmetric, cold and lukewarm black holes in cosmological Einstein--Maxwell theory,''
Nucl.\ Phys.\ B \textbf{383}, 395 (1992), arXiv:hep-th/9203018.

\bibitem{Smarr1973}
L.~Smarr,
``Mass formula for Kerr black holes,''
Phys.\ Rev.\ Lett.\ \textbf{30}, 71 (1973); Erratum: \textbf{30}, 521 (1973).

\bibitem{KRT2009}
D.~Kastor, S.~Ray, J.~Traschen,
``Enthalpy and the mechanics of AdS black holes,''
Class.\ Quantum Grav.\ \textbf{26}, 195011 (2009), arXiv:0904.2765.

\bibitem{Nariai1950}
H.~Nariai,
``On some static solutions of Einstein's gravitational field equations in a spherically symmetric case,''
Sci.\ Rep.\ T\^ohoku Univ.\ Ser.\ I \textbf{34}, 160 (1950); reprinted in Gen.\ Rel.\ Grav.\ \textbf{31}, 951 (1999), doi:10.1023/A:1026698508110.

\bibitem{Nariai1951}
H.~Nariai,
``On a new cosmological solution of Einstein's field equations of gravitation,''
Sci.\ Rep.\ T\^ohoku Univ.\ Ser.\ I \textbf{35}, 46 (1951); reprinted in Gen.\ Rel.\ Grav.\ \textbf{31}, 963 (1999).

\bibitem{GP1983}
P.~Ginsparg, M.~J.~Perry,
``Semiclassical perdurance of de Sitter space,''
Nucl.\ Phys.\ B \textbf{222}, 245 (1983).

\bibitem{Harada2023}
J.~Harada,
``Gravity at cosmological distances: explaining the accelerating expansion without dark energy,''
Phys.\ Rev.\ D \textbf{108}, 044031 (2023), arXiv:2308.02115.

\bibitem{MM2023}
C.~A.~Mantica, L.~G.~Molinari,
``Note on Harada's conformal Killing gravity,''
Phys.\ Rev.\ D \textbf{108}, 124029 (2023), arXiv:2308.06803.

\bibitem{MM2024}
C.~A.~Mantica, L.~G.~Molinari,
``Conformal Killing gravity in static spherically-symmetric spacetimes,''
Phys.\ Rev.\ D \textbf{110}, 044025 (2024), arXiv:2406.12511.

\bibitem{CN2024}
G.~Cl\'ement, K.~Nouicer,
``Spherical symmetric solutions of conformal Killing gravity: black holes, wormholes, and sourceless cosmologies,''
Class.\ Quantum Grav.\ \textbf{41}, 165005 (2024), arXiv:2404.00328.

\bibitem{Barnes2023}
A.~Barnes,
``Harada--Maxwell static spherically symmetric spacetimes,''
arXiv:2311.09171 (2023).

\bibitem{Akhtar2023}
Z.~Akhtar, A.~Khan, Z.~Ahmad, R.~Ali,
``Study of first-order corrections and phase transition of RN de-Sitter black hole,''
New Astron.\ \textbf{98}, 101936 (2023), doi:10.1016/j.newast.2022.101936. 

\bibitem{Casana2018}
R.~Casana, A.~Cavalcante, F.~P.~Poulis, E.~B.~Santos,
``Exact Schwarzschild-like solution in a bumblebee gravity model,''
Phys.\ Rev.\ D \textbf{97}, 104001 (2018), arXiv:1711.02273.

\bibitem{Liu2025}
J.-Z.~Liu, W.-D.~Guo, S.-W.~Wei, Y.-X.~Liu,
``Charged spherically symmetric and slowly rotating charged black hole solutions in bumblebee gravity,''
Eur.\ Phys.\ J.\ C \textbf{85}, 145 (2025), arXiv:2407.08396.

\end{thebibliography}
\end{document}